\documentclass[usenatbib]{mn2e}

\usepackage{amsmath}
\usepackage{natbib}
\usepackage{epsfig}
\usepackage{txfonts}
\usepackage{pict2e}
\usepackage[usenames,dvipsnames]{color}

\newcommand{\Mpc}{{\rm Mpc}}

\newcommand{\expf}[1]{{{\rm e}^{#1}}}

\newcommand{\id}{{\,\rm d}}

\newcommand{\beq}{\begin{equation}}   %

\newcommand{\eeq}{\end{equation}}   %

\newcommand{\beqa}{\begin{eqnarray}}   %

\newcommand{\eeqa}{\end{eqnarray}}   %

\newcommand{\beal}{\begin{align}}
\newcommand{\enal}{\end{align}}

\newcommand{\bspl}{\begin{split}}

\newcommand{\espl}{\end{split}}

\newcommand{\bsub}{\begin{subequations}}

\newcommand{\esub}{\end{subequations}}

\newcommand{\bmulti}{\begin{multline}}   %

\newcommand{\beqm}{\begin{mathletters}}   %

\newcommand{\eeqm}{\end{mathletters}}   %

\newcommand{\me}{m_{\rm e}}

\newcommand{\Ne}{N_{\rm e}}

\newcommand{\Te}{T_{\rm e}}

\newcommand{\Tg}{T_{\gamma}}

\newcommand{\sigT}{\sigma_{\rm T}}

\newcommand{\aEMs}{\alpha_{\rm EM, 0}}
\newcommand{\mes}{m_{\rm e, 0}}
\newcommand{\aEM}{\alpha_{\rm EM}}
\newcommand{\dalpha}{\Delta\alpha/\alpha}
\newcommand{\dme}{\Delta m_{\rm e}/m_{\rm e}}
\newcommand{\LCDM}{$\Lambda$CDM}
\newcommand{\changeJ}[1]{{\textcolor{black}{#1}}}

\usepackage{hyperref}
\usepackage{grffile}
\usepackage{graphics}
\usepackage{booktabs}

\title[Variation of $\aEM$ and $\me$]
{New constraints on time-dependent variations of fundamental constants using {\it Planck} data}

\author[L. Hart and J. Chluba]{
Luke Hart$^{1}$\thanks{Email: luke.hart@postgrad.manchester.ac.uk}
and Jens Chluba$^{1}$\thanks{Email: jens.chluba@manchester.ac.uk}
\\
$^{1}$Jodrell Bank Centre for Astrophysics, Alan Turing Building, University of Manchester, Manchester M13 9PL
}

\date{\vspace{-3mm}Accepted 2017 --. Received 2017 May 10.}

\pubyear{2017}

\begin{document}
\label{firstpage}
\pagerange{\pageref{firstpage}--\pageref{lastpage}}
\maketitle
\begin{abstract}
Observations of the cosmic microwave background (CMB) today allow us to answer detailed questions about the properties of our Universe, targeting both {\it standard} and {\it non-standard} physics.
In this paper, we study the effects of varying fundamental constants (i.e., the fine-structure constant, $\aEM$, and electron rest mass, $\me$) around last scattering using the recombination codes {\tt CosmoRec} and {\tt Recfast++}. We approach the problem in a pedagogical manner, illustrating the importance of various effects on the free electron fraction, Thomson visibility function and CMB power spectra, highlighting various degeneracies. We demonstrate that the simpler {\tt Recfast++} treatment (based on a three-level atom approach) can be used to accurately represent the full computation of {\tt CosmoRec}. We also include {\it explicit} time-dependent variations using a phenomenological power-law description. We reproduce previous {\it Planck} 2013 results in our analysis. Assuming constant variations relative to the standard values, we find the improved constraints $\aEM/\aEMs=0.9993\pm 0.0025$ (CMB only) and $\me/\mes= 1.0039 \pm 0.0074$ (including BAO) using {\it Planck} 2015 data. For a redshift-dependent variation, $\aEM(z)=\aEM(z_0)\,[(1+z)/1100]^p$ with $\aEM(z_0)\equiv\aEMs$ at $z_0=1100$, we obtain $p=0.0008\pm 0.0025$. Allowing simultaneous variations of $\aEM(z_0)$ and $p$ yields $\aEM(z_0)/\aEMs = 0.9998\pm 0.0036$ and $p = 0.0006\pm 0.0036$. We also discuss combined limits on $\aEM$ and $\me$. Our analysis shows that existing data is not only sensitive to the value of the fundamental constants around recombination but also its first time derivative. This suggests that a wider class of varying fundamental constant models can be probed using the CMB.
\end{abstract}

\begin{keywords}
recombination -- fundamental physics -- cosmology -- CMB anisotropies
\end{keywords}

\section{Introduction}
\label{sec:intro}
Nowadays, measurements of the cosmic microwave background (CMB) anisotropies allow us to constrain the standard cosmological parameters with unprecedented precision \citep{wmap9results, Planck2015params}. This has opened a route for testing possible extensions to the \LCDM~model, e.g., related to the effective number of neutrino species and their mass \citep[see][]{PlanckNeutrino, BattyeNeutrinos, Abazajian2015} and Big Bang Nucleosynthesis \citep{Coc2013, PRISM2013WPII, CMBS42016}. In the analysis, we are furthermore sensitive to percent level effects in the recombination dynamics \citep{RubinoMartin2010, Shaw2011}, which can be captured using advanced recombination codes such as {\tt CosmoRec} and {\tt HyRec} \citep{Chluba2010b,Yacine2010c}, again emphasizing the impressive precision of available cosmological datasets.

Our interpretation of the CMB measurements relies on several assumptions. The validity of general relativity and atomic physics around recombination are two evident ones. This encompasses a significant extrapolation of local physics, tested in the lab, to cosmological scales (both in distance and time). Albeit the successes of the \LCDM~cosmology, we know that simple extrapolation is currently not enough to explain the existence of dark matter and dark energy in our Universe. Similarly, it is important to test the validity of local physical laws in different regimes. 

One of these tests is related to the {\it constancy of fundamental constants} \citep[see][for review]{Uzan2003,Uzan2011}. This could provide a glimpse at physics beyond the standard model, possibly shedding light on the presences of additional scalar fields and their coupling to the standard sector. Variations of the fine-structure constant, $\alpha_{\rm EM}$, and electron rest mass, $m_{\rm e}$ can directly impact CMB observables, as studied previously \citep[e.g.,][]{Kaplinghat1999, Avelino2000, Battye2001, Avelino2001, Rocha2004, Martins2004, Scoccola2009, Menegoni2012} using modified versions of {\tt recfast} \citep{SeagerRecfast1999}.
Similarly, changes of the gravitational constant can be considered \citep[e.g.,][]{Aguilar2003,Martins2010, Galli2011}.
Using {\it Planck} 2013 data, the values of $\aEM$ and $\me$ around recombination were proven to coincide with those obtained in the lab to within $\simeq 0.4\%$ for $\aEM$ and $\simeq 1-6\%$ for $\me$ \citep{Planck2015var_alp}. This is $\simeq 2-3$ orders of magnitude weaker than constraints obtained from `local' measurements \citep{Bize2003,Rosenband2008, Bonifacio2014, Kotus2017}; however, the CMB places limits during very different phases in the history of the Universe, some $380,000$ years after the Big Bang, which complement these low-redshift measurements.

In this paper, we describe the effects of varying fundamental constants on the cosmological recombination history, focusing on variations of $\aEM$ and $\me$. These directly affect the atomic physics and  radiative transfer in the recombination era ($z\simeq 10^3$) and thus can be probed using CMB anisotropy measurements. We approach the problem is a pedagogical manner, illustrating the individual effects on the recombination dynamics in Sec.~\ref{section:effects}. We show that the full recombination calculation of {\tt CosmoRec} can be accurately represented using a simple three-level atom approach, by introducing appropriate corrections functions (see Sect.~\ref{sec:mod_Recfast++}). We discuss constant changes of $\aEM$ and $\me$, but also introduce a phenomenological power-law redshift-dependence of these parameters around recombination. The associated effects on the ionization history are distinct and thus can be individually constrained using CMB data.

The changes to the recombination codes are then propagated to the Thomson visibility function and the calculations of the CMB power spectra. Here we do not focus on the individual contributions to the CMB power spectrum deviations as these have been covered in previous literature \citep[see][Appendix B]{Planck2015var_alp}. However, we illustrate the overall effects and also highlight existing degeneracies between changes caused by $\aEM$, $\me$ and the average CMB temperature, $T_0$. In Sect.~\ref{sec:constraints}, we present our constraints for different cases using {\it Planck} 2015 data. In particular, we find the CMB data to be sensitive not only to the value of the fundamental constants around recombination but also its first time derivative. We conclude in Sect.~\ref{sec:conc}.

\vspace{-4mm}
\section{Effects of varying fundamental constants on ionization history}
\label{sec:rec_phys_var}
In this section, we describe the effects of varying $\alpha_{\rm EM}$ and $m_{\rm e}$ on the ionization history. We use modified versions of {\tt CosmoRec} and {\tt Recfast++} \citep{Chluba2010} for our computations\footnote{These codes are available at \url{www.Chluba.de/CosmoRec}.}, highlighting the importance of different effects and their individual impact on the free electron fraction, $X_{\rm e}$. 

\vspace{-4mm}
\subsection{How do $\alpha_{\rm EM}$ and $\me$ enter the recombination problem?}
\label{section:effects}
Varying $\alpha_{\rm EM}$ and $m_{\rm e}$ inevitably creates changes in the ionization history. Most importantly, the energy levels of hydrogen and helium depend on these constants, $E_i\propto\alpha_{\rm EM}^2m_{\rm e}$, which directly affects the recombination redshift. In addition, atomic transition rates and photoionization/recombination rates are altered. Lastly, the interactions of photons and electrons through Compton and resonance scattering modify the radiative transfer physics. 
In an effective three-level atom approach \citep{Zeldovich68, Peebles68, Seager2000}, the individual dependencies can be summarized as \citep[e.g.,][]{Kaplinghat1999, Scoccola2009, Planck2015var_alp, Chluba2016CosmoSpec}
\begin{align}
\label{eq:recfast_scaling}
\begin{split}
\sigma_{\rm T} \propto \alpha_{\rm EM}^2 m_{\rm e}^{-2} 
\qquad A_{2\gamma} &\propto \alpha_{\rm EM}^8 m_{\rm e} 
\qquad P_{\rm S} A_{1\gamma} \propto \alpha_{\rm EM}^{6}m_{\rm e}^{3} 
\\
\alpha_{\rm rec} \propto \alpha_{\rm EM}^2 m_{\rm e}^{-2} 
\qquad \beta_{\rm phot} &\propto \alpha_{\rm EM}^5 m_{\rm e} 
\qquad T_{\rm eff} \propto \alpha_{\rm EM}^{-2}m_{\rm e}^{-1}.
\end{split}
\end{align} 
Here, $\sigma_{\rm T}$ denotes the Thomson scattering cross section; $A_{2\gamma}$ is the two-photon decay rate of the second shell; $\alpha_{\rm rec}$ and $\beta_{\rm phot}$ are the effective recombination and photoionization rates, respectively; $T_{\rm eff}$ is the effective temperature at which $\alpha_{\rm rec}$ and $\beta_{\rm phot}$ need to be evaluated (see explanation below); $P_{\rm S}A_{1\gamma}$ denotes the effective dipole transition rate for the main resonances (e.g., Lyman-$\alpha$), which is reduced by the Sobolev escape probability, $P_{\rm S}\leq 1$ \citep{Sobolev1960, Seager2000} with respect to the vacuum rate, $A_{1\gamma}$.
For a more detailed account of how the transition rates depend on the fundamental constants we refer to \citet{Chluba2016CosmoSpec} and the manual of {\tt HyRec} \citep{Yacine2010c}. 

The scalings of $\sigma_{\rm T}$, $A_{2\gamma}$ and $P_{\rm S}A_{1\gamma}$ directly follow from their explicit dependencies on $\alpha_{\rm EM}$ and $m_{\rm e}$.  The shown scalings of $\alpha_{\rm rec}$ and $\beta_{\rm phot}$ reflect renormalisations of the transition rates, again stemming from their explicit dependencies on $\alpha_{\rm EM}$ and $m_{\rm e}$ \citep[e.g.,][]{Karzas1961}. However, these rates also depend directly on the ratio of the electron/photon temperature to the ionization threshold. This leads to an additional dependence on $\alpha_{\rm EM}$ and $m_{\rm e}$, which can be captured by evaluating these rates at rescaled temperature, with scaling indicated through $T_{\rm eff}$.
Overall, this leads to the effective dependence $\alpha_{\rm rec}\propto \alpha_{\rm EM}^{3.44} \, \me^{-1.28}$ around hydrogen recombination  \citep{Chluba2016CosmoSpec}. The required photoionization rate, $\beta_{\rm phot}$, is obtained using the detailed balance relation. \changeJ{Slightly different overall scalings for $\alpha_{\rm rec}$ and $\beta_{\rm phot}$ were used in \cite{Planck2015var_alp}, but we find the associated effect on the recombination history to be sub-dominant and limited to $z\lesssim800$.}

We also highlight that all atomic species are treated using hydrogenic scalings. For neutral helium, non-hydrogenic effects (e.g., fine-structure transitions, singlet-triplet couplings) become relevant \citep{DrakeBook2006}. However, the corrections should be sub-dominant and are neglected here.

We will illustrate the importance of the different terms in Eq.~\eqref{eq:recfast_scaling} in Sect.~\ref{sec:effect_Xe}. This will show that in particular the changes in the energy scale, which are captured by rescaling the temperature, are crucial. We now continue by explaining the required modifications to {\tt Recfast++} and {\tt CosmoRec}.

\begin{figure}
	\includegraphics[width=\columnwidth]{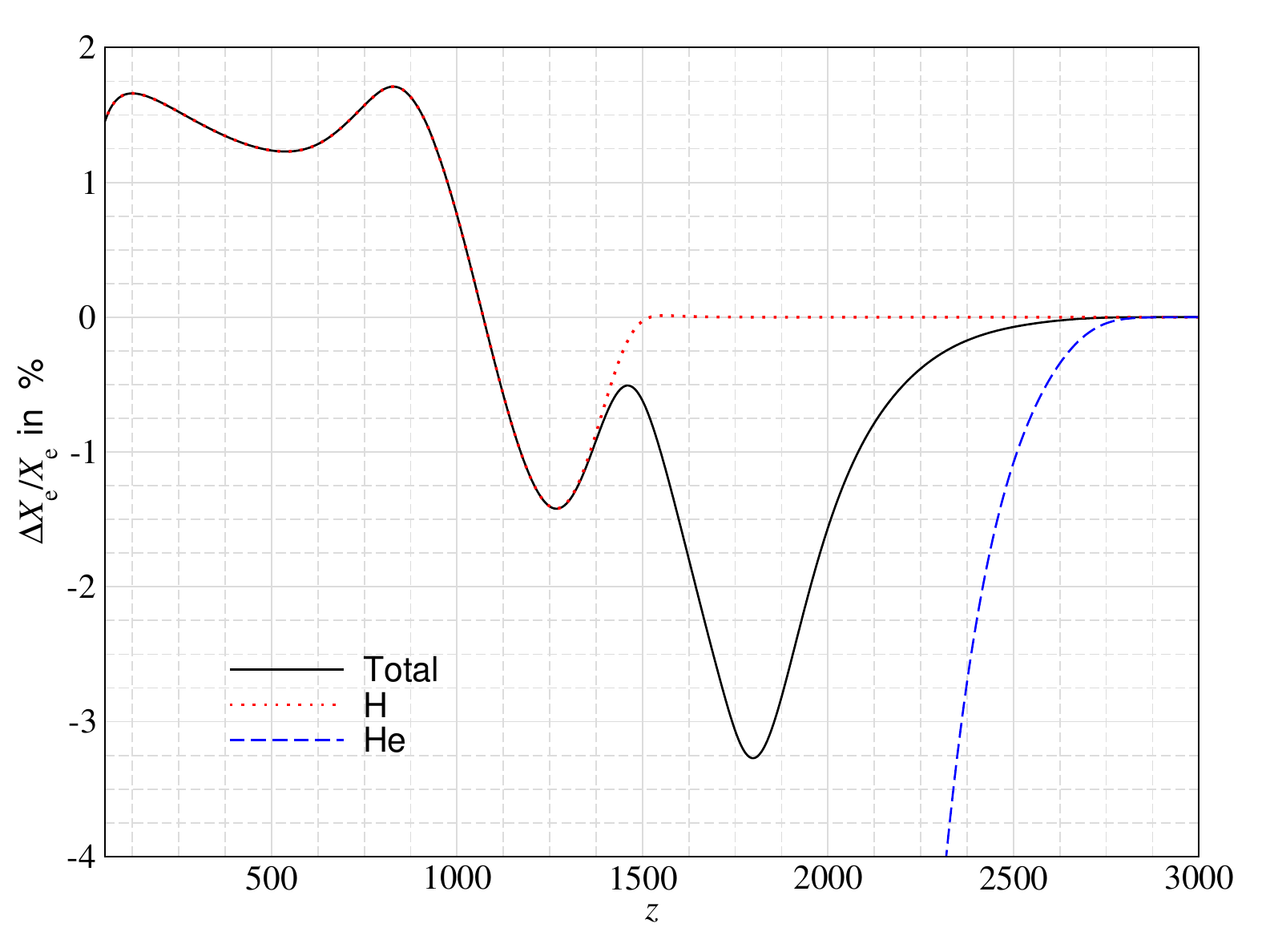}
    \caption{Relative difference in the free electron fractions of {\tt CosmoRec} and {\tt Recfast++} ({\tt Recfast++} is used as reference) for the standard cosmology. The lines show the results for \changeJ{$X_{\rm e}=X^{\rm H}_{\rm e}+X^{\rm He}_{\rm e}$} (black/solid), $X^{\rm H}_{\rm e}$ (dotted/red) and $X^{\rm He}_{\rm e}$ (dashed/blue).}
    \label{fig:correction}
\end{figure}

\vspace{-2mm}
\subsubsection{Modifications to \texttt{Recfast++}}
\label{sec:mod_Recfast++}
{\tt Recfast++} is based on a simple three-level atom approach, similar to that of {\tt recfast} \citep{SeagerRecfast1999}. It evaluates three ordinary differential equations, evolving the free electron fraction contributions from hydrogen and singly-ionized helium, $X_{\rm e}^{\rm H}$ and $X_{\rm e}^{\rm He}$, respectively, as well as the matter/electron temperature, $\Te$. Doubly-ionized helium is modeled using the Saha-relations. 

As an added feature of the \texttt{Recfast++} code, one can modify the obtained ionization history with a \textit{correction function} to represent the full recombination calculation of \texttt{CosmoRec} \citep{Jose2010, Shaw2011}.  The required correction function between \texttt{Recfast++} and \texttt{CosmoRec} is obtained as
 \begin{equation}
 \label{eq:CF_tot}
X_{\rm e}^{\rm C}(z) \approx \left(1+ \frac{\Delta X_{\rm e}(z)}{X_{\rm e}^{\rm R}(z)}\right)\,X_{\rm e}^{\rm R}(z)=f_{\rm tot}(z)\,X_{\rm e}^{\rm R}(z),
 \end{equation}
where '${\rm C}$' refers to \texttt{CosmoRec}, '${\rm R}$' to \texttt{Recfast++}. In the code, the relative difference, $\Delta X_{\rm e}/X_{\rm e}^{\rm R}=(X_{\rm e}^{\rm C}-X_{\rm e}^{\rm R})/X_{\rm e}^{\rm R}$, is stored for the standard cosmology and then interpolated to obtain $f_{\rm tot}(z)$. The relative difference, $\Delta X_{\rm e}/X_{\rm e}^{\rm R}$, is illustrated in Fig.~\ref{fig:correction}.
For the standard cosmology, {\tt Recfast++} naturally allows a quasi-exact representation of the full calculation. For small variations around the standard cosmology, the correction-to-correction can be neglected so that this approach remains accurate in CMB analysis \citep{Shaw2011, Planck2015params}. 

All the modifications listed in Eq.~\eqref{eq:recfast_scaling} are readily incorporated to the simple {\tt Recfast++} equations. However, we found that for our purpose it was beneficial to treat the correction functions for hydrogen and helium separately, since in the transition regime between hydrogen and helium recombination ($z \simeq 1600$) the free electron fraction departs from unity, which is physically not expected. This generalizes Eq.~\eqref{eq:CF_tot} to
\begin{equation}
X^{\rm C}_{\rm e}(z) \approx f_{\rm H}(z)\,X_{\rm e}^{\rm H, R}(z)+f_{\rm He}(z)\,X_{\rm e}^{\rm He, R}(z),
\end{equation}
where we multiply each correction function term with its respective contribution to the total $X_{\rm e}=X_{\rm e}^{\rm H}+X_{\rm e}^{\rm He}$. The individual correction functions are again obtained using relative differences with respect to the standard cosmology, $f_{i}(z)=1+\Delta X_{\rm e}^{i}/X_{\rm e}^{i, \rm R}$.
This is illustrated in Figure \ref{fig:correction}. At $z\simeq 2500$, the helium correction sharply drops to $\Delta X_{\rm e}^{\rm He}/X_{\rm e}^{\rm He, R}\approx$ \changeJ{$-100\%$ (i.e. $f_{\rm He} \rightarrow 0$)}, indicating that helium rapidly recombines. This is related to hydrogen continuum absorption of helium photons, which is not taken into account in the standard treatment \citep{Kholupenko2007, Switzer2007I, Jose2008}. Since hydrogen recombination occurs at lower redshifts, the hydrogen corrections tend to $0$ at $z\gtrsim 1500$, while around $z\simeq1100$ radiative transfer corrections become visible \citep[e.g.,][for overview]{Fendt2009, Jose2010}. 

At $z\lesssim 1500$, $f_{\rm tot}(z)\approx f_{\rm H}(z)$, while the features related to helium recombination corrections around $z\simeq 1700$ are now represented directly by the helium correction function. Once added to {\tt Recfast++}, it more fairly weights the helium corrections than the previous approach. In the code, one can chose between the two versions, but we find that when varying the fundamental constants, the new approach works best. It is furthermore important to interpret the correction functions as a function of temperature. This leads to the remapping $z\rightarrow z \times(\aEM/\aEMs)^{-2}(\me/\mes)^{-1}$, which captures the leading order transformation of radiative transfer corrections.

\vspace{-0mm}
\subsubsection{Modifications to \texttt{CosmoRec}}
\label{sec:mod_CosmoRec}
The modifications to {\tt Recfast++} were relatively straightforward. However, for {\tt CosmoRec} this became a slightly bigger task.
{\tt CosmoRec} is built up as a modular system that allows each {\it module} to act as a plugin. In {\tt CosmoRec}, the energies and transition rates within the hydrogen and neutral helium atoms needed to be rescaled with the previously mentioned scalings. These are represented by classes called {\tt Atom} and {\tt HeI\_Atom}, respectively, which include all the properties of given atomic levels, the collection of levels that form the atom and the ensemble of atoms around recombination. These can also be used as independent coding modules for atomic physics calculations. The neutral helium scalings with $\aEM$ and $\me$ are modeled using hydrogenic expressions, which is expected to be accurate at the $\simeq 0.1\%-1\%$ level but omits higher order effects to the energy levels or transition rates.  

After the atomic initializations, the effective transition rates \citep[see][for details about the method]{Yacine2010} related to the multi-level atom need to be rescaled. In the code, these affect the effective recombination rates, $\mathcal{A}(T_\gamma,T_{\rm e})$, the photoionization rates, $\mathcal{B}(T_\gamma)$ and the inter-state transition rates, $\mathcal{R}(T_\gamma)$. Changes related to $\sigT$ are again trivial to include.

During recombination, the processes occurring within the atoms are influenced by the temporal evolution of the background photon field. This complicates the recombination problem with the need for partial differential equations (PDEs) describing the radiative transfer \citep[e.g.][]{Chluba2007b, Grachev2008, Chluba2008b, Hirata2008, Hirata2009, Chluba2009b}. When the fundamental constants are modified, one must again rescale the rates and energies required for the computations of the photon field. Similarly, the two-photon and Raman scattering profiles \citep{Chluba2008a, Hirata2008, Chluba2010b} have to be altered. We also carefully considered modifications to the neutral helium radiative transfer \citep{Chluba2012HeRec}. These effects can be separately activated in the latest version of {\tt CosmoRec} (i.e. version 3.0 or higher).

\begin{figure}
	\includegraphics[width=\columnwidth]{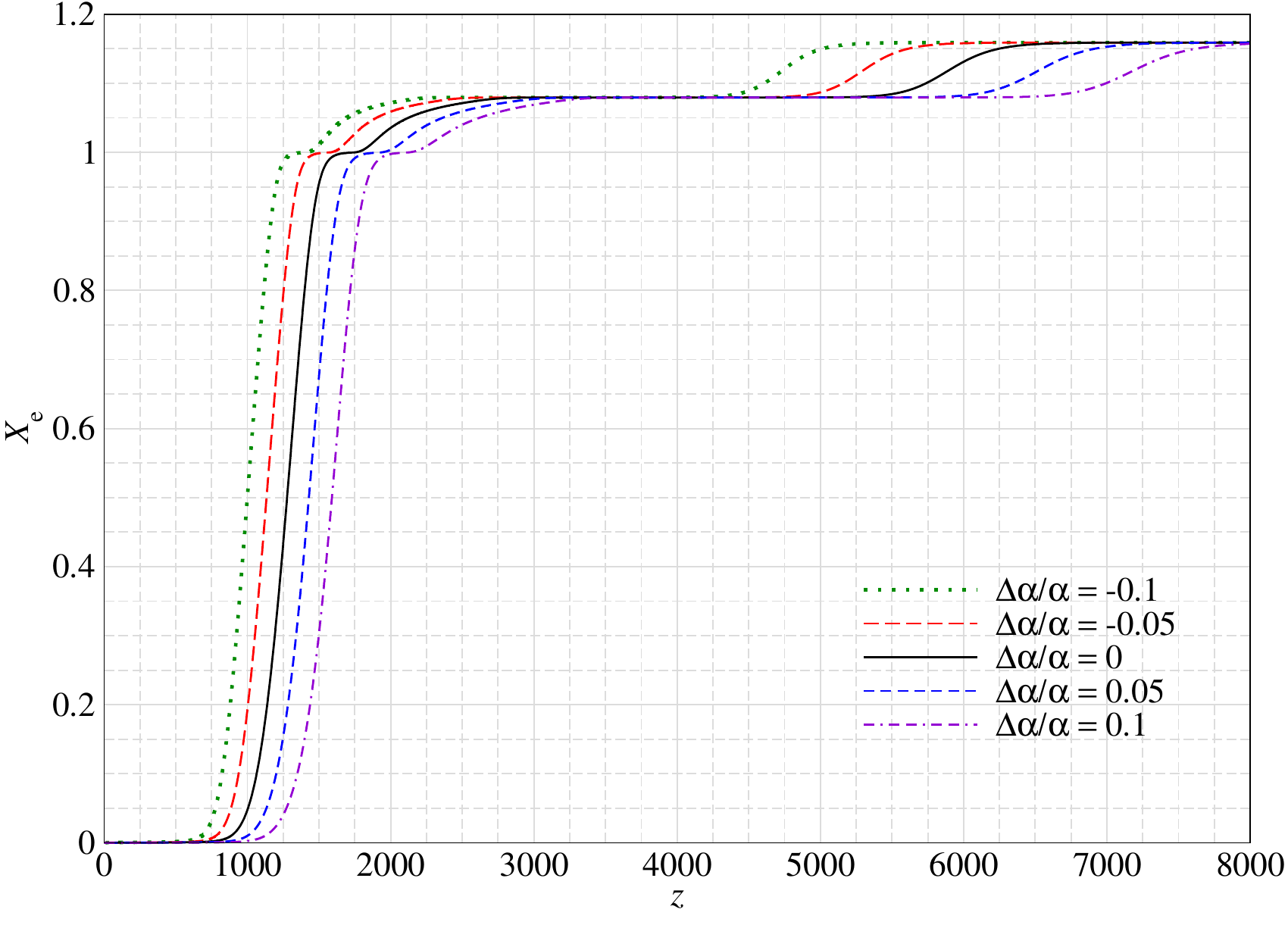}
    \caption{Ionization histories for different values of $\aEM$. The dominant effect is caused by modifications of the ionization threshold, which implies that for increased $\aEM$ recombination finishes earlier. The curves were computed using {\tt Recfast++}.}
    \label{fig:alpha}
\end{figure}

\vspace{2mm}
\subsection{Relevance of different effects for $X_{\rm e}$}
\label{sec:effect_Xe}
We now illustrate the importance of the individual effects in Eq.~\eqref{eq:recfast_scaling}, for now assuming {\it constant} changes of $\aEM$ and $\me$. This will be generalized in Sect.~\ref{sec:time_dep_alpha}. We shall start by focusing on changes caused by varying $\aEM$, parametrized as $\aEM=\aEMs(1+\dalpha)$.
When all the terms relevant to the recombination problem are included, we obtain the ionization histories shown in Fig.~\ref{fig:alpha} for different values of $\dalpha$. Increasing the fine structure constant shifts the moment of recombination toward higher redshifts. This agrees with the results found earlier in \citet{Kaplinghat1999}, \citet{Battye2001} and \citet{Rocha2004} and can intuitively be understood in the following manner: $\dalpha>0$ increases the transition energies between different atomic levels and the continuum. This increases the energy threshold at which recombination occurs, hence increasing the recombination redshift, an effect that is basically captured by an effective temperature rescaling in the evaluation of the photoionization and recombination rates (see below). 

\begin{figure}
	\includegraphics[width=\columnwidth]{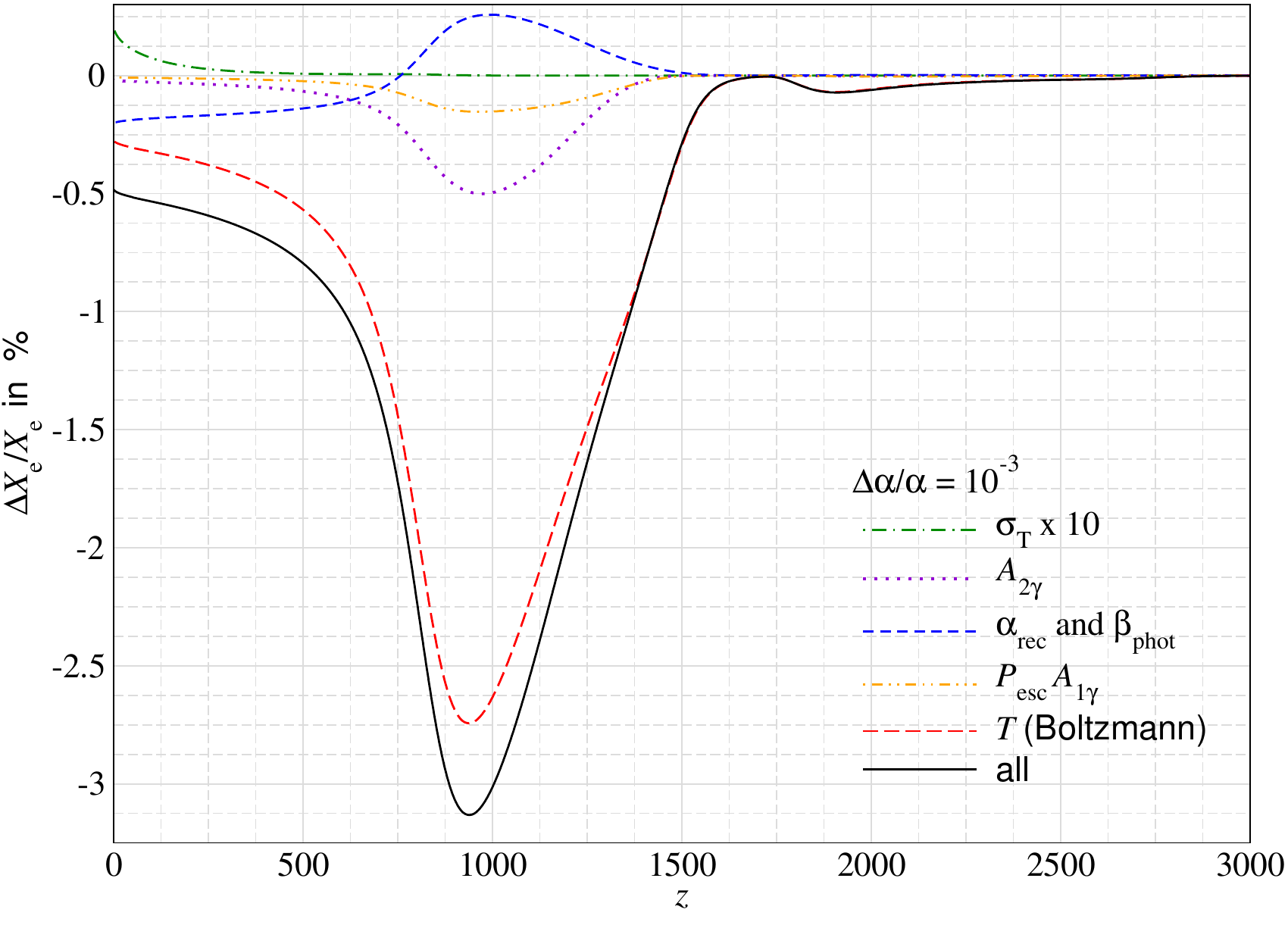}
    \caption{The relative changes in the ionization history for $\dalpha=10^{-3}$ with respect to the standard case caused by different effects. {\tt Recfast++} was used for the computations. The rescaling of temperature ($\leftrightarrow$ mainly affecting the Boltzmann factors) yields \changeJ{$\Delta X_{\rm e}/X_{\rm e}\simeq -2.7\%$}, dominating the total contributions, which peaks with $\simeq -3.1\%$ at $z\simeq 1000$. Note that the modification due to $\sigma_{\rm T}$ has been scaled by $10$ to make it visible.}
    \label{fig:cont}
\end{figure}
The relative changes to the ionization history, $\Delta X_{\rm e}/X_{\rm e}$, for the different terms discussed in Section \ref{section:effects} are illustrated in Fig.~\ref{fig:cont}. We chose a value for $\dalpha=10^{-3}$, which leads to a percent-level effect on $X_{\rm e}$.
As expected, the biggest effect appears after rescaling the temperature for the evaluation of the photonionization and recombination rates. More explicitly, this can be understood when considering the net recombination rate to the second shell, which can be written as $\Delta R_{\rm con}= \Ne N_{\rm p} \alpha_{\rm rec}-N_{\rm 2}\,\beta_{\rm phot}=\alpha_{\rm rec}[\Ne N_{\rm p}-g(\Tg)N_{\rm 2}]$ (in full equilibrium, $\Delta R_{\rm con}=0$), where $g(\Tg)\propto \Tg^{3/2}\,\expf{-h\nu_{\rm 2c}/k\Tg}$ with continuum threshold energy, $E_{\rm 2c}=h\nu_{\rm 2c}$.
Here, the exponential factor ($\leftrightarrow$ Boltzmann factor) is most important, leading to an enhanced effect once the replacement $\Tg'(z)=\Tg(z) \times(\aEM/\aEMs)^{-2}(\me/\mes)^{-1}$ is carried out. For $\dalpha=10^{-3}$, this gives $\Delta X_{\rm e}/X_{\rm e}\simeq -2.7\%$ at $z\simeq 1000$, which accounts for nearly all of the effect (cf., Fig.~\ref{fig:cont}). 

The second largest individual effect is due to the rescaling of the two-photon decay rate, $A_{2\gamma}$. This is expected since $\aEM$ appears in a high power, $A_{2\gamma}\propto \aEM^8$, and because the 2s-1s two-photon channel plays such a crucial role for the recombination dynamics \citep{Zeldovich68, Peebles68, Chluba2006}, allowing $\simeq 58\%$ of all hydrogen atoms to become neutral through this route \citep{Chluba2006b}. For $\dalpha=10^{-3}$, we find $\Delta X_{\rm e}/X_{\rm e}\simeq -0.5\%$ at $z\simeq 1000$. 

The normalizations of the recombination and photoionization rates (blue/dashed line) give rise to a net delay of $\Delta X_{\rm e}/X_{\rm e}\simeq 0.3\%$ at $z\simeq1000$, which partially cancels the correction due to $A_{2\gamma}$. This is due to the stronger scaling of $\beta_{\rm phot}$ with $\aEM$ than $\alpha_{\rm rec}$. At low redshifts ($z\lesssim 750$), recombination is again accelerated, indicating that a higher fraction of recombination events occurs, as the importance of photoionization ceases. 
The correction related to the Lyman-$\alpha$ channel is found to be $\simeq 3.3$ times smaller than for the two-photon channel, yielding $\Delta X_{\rm e}/X_{\rm e}\simeq -0.15\%$ at $z\simeq1000$ (cf., Fig.~\ref{fig:cont}).

Figure~\ref{fig:cont} also shows that the contributions from rescaling $\sigma_{\rm T}$ are very small and only become noticeable at low redshifts. At these redshifts, the matter and radiation temperature begins to depart, giving $\Te < \Tg$. For larger $\aEM$, this departure is delayed, such that $\Te$ stays longer close to $\Tg$. Hotter electrons recombine less efficiently, so that a slight delay of recombination appears (cf., Fig.~\ref{fig:cont}).
We find that this correction can in principle be neglected without affecting the results notably, but include it for completeness.

\begin{figure}
  \includegraphics[width=\columnwidth]{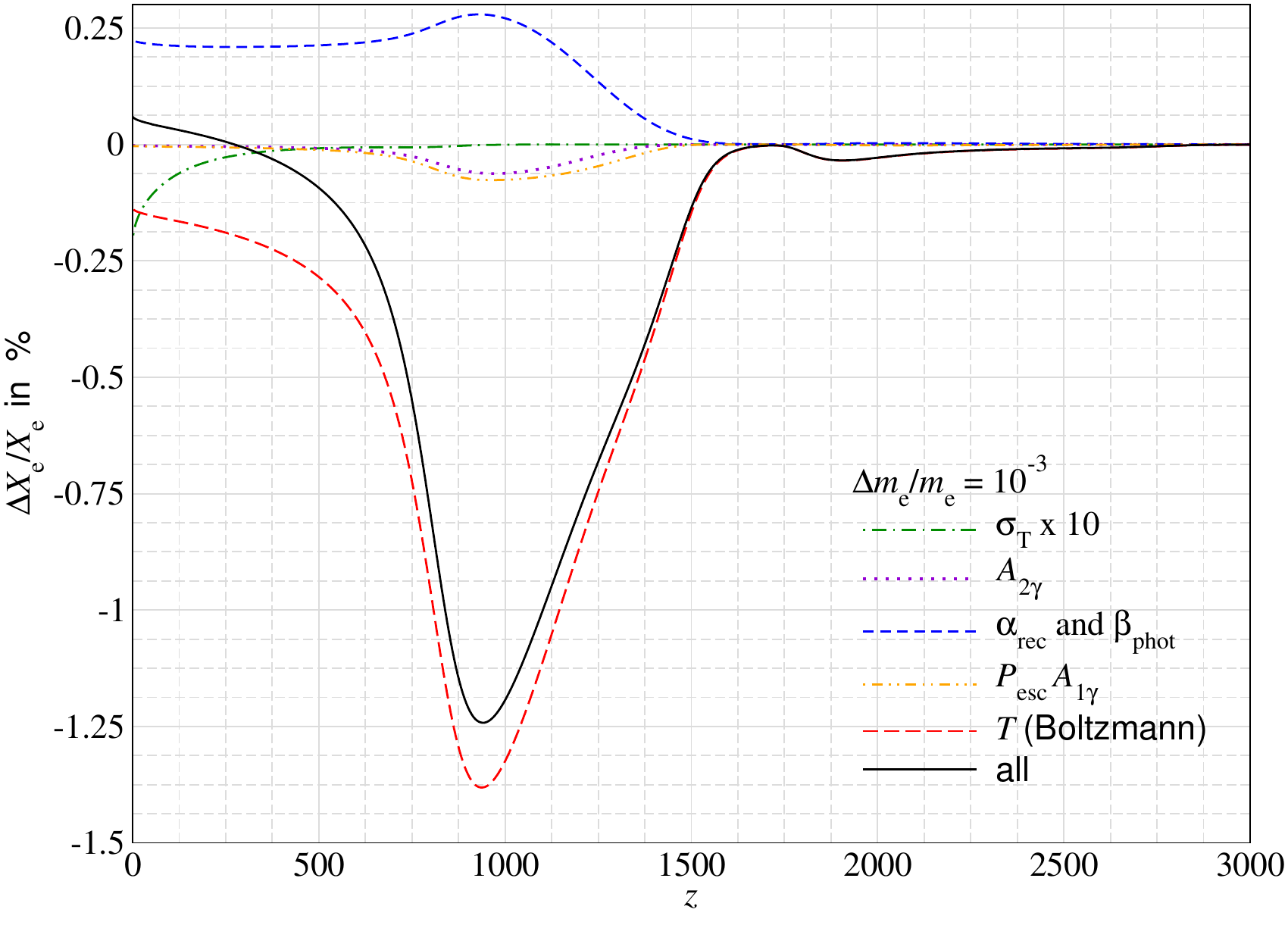}
  \caption{Same as in Fig.~\ref{fig:cont} but for $\dme=10^{-3}$. The effective temperature rescaling again dominates the total change. Around $z\simeq 1000$, the total effect is $\simeq 2.5$ times smaller than for $\dalpha=10^{-3}$.}
  \label{fig:cont_me}
\end{figure}

\begin{figure}
	\includegraphics[width=\columnwidth]{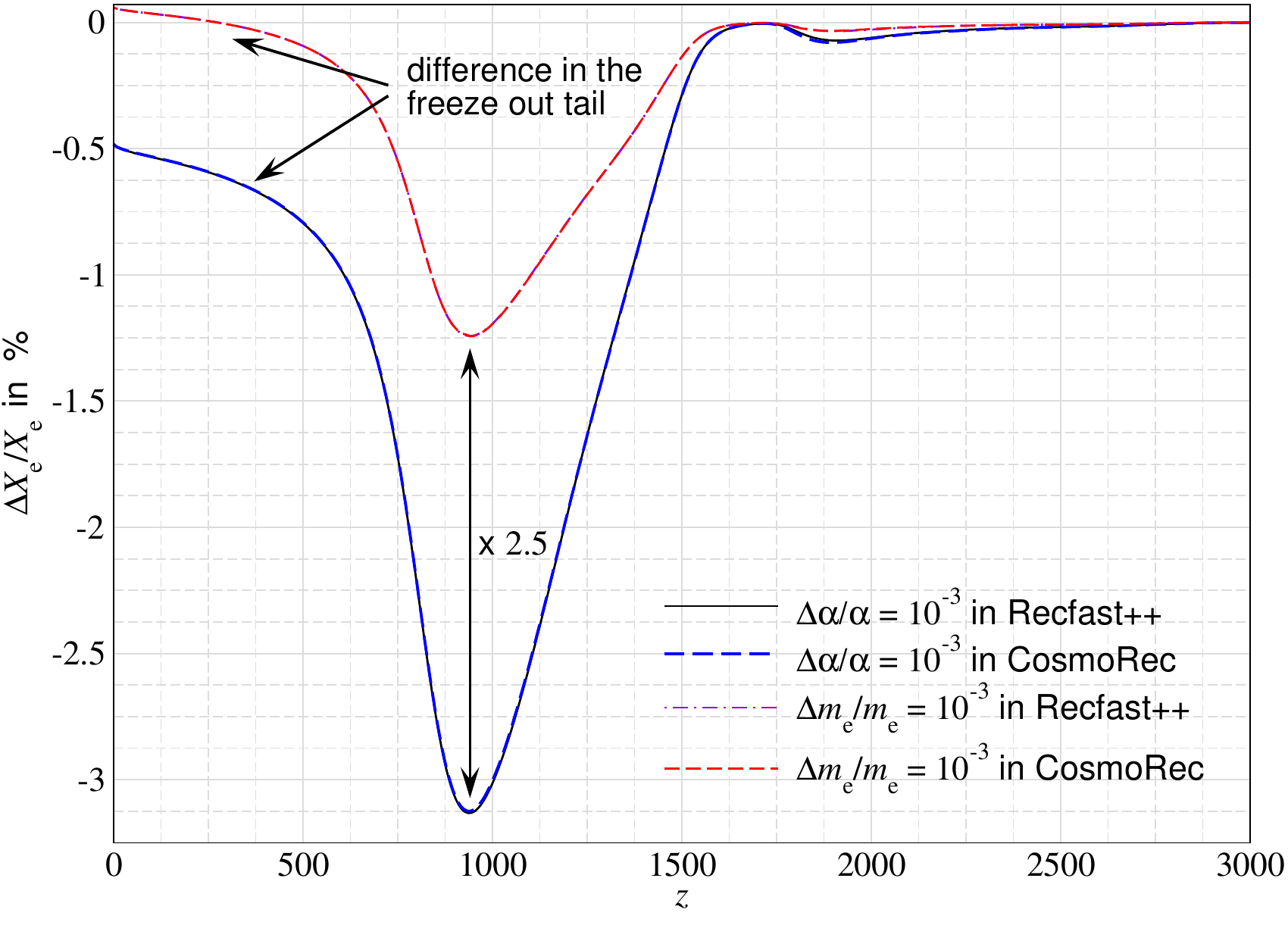}
    \caption{Comparison of the changes to the ionization history caused by variation of $\aEM$ and $\me$ as computed with {\tt Recfast++} and {\tt CosmoRec}. Both codes agree extremely well (the lines practically overlap for the two cases), departing only at the level of $\simeq 0.1\%$ for $X_{\rm e}$. For changes of $\me$, the effect on the freeze-out tail is much smaller than for $\aEM$.}
    \label{fig:compare}
\end{figure}

\vspace{-0mm}
\subsubsection{Changes due to variation of $\me$}
\label{sec:Xe_me}
We now focus on changes caused by the effective electron mass, parametrized as $\me=\mes(1+\dme)$. Inspecting the scalings of Eq.~\eqref{eq:recfast_scaling}, we expect the overall effect to be smaller than for $\aEM$. For example, the effect of temperature rescaling should be roughly half as large. Similarly, the effect due to rescaling $A_{2\gamma}$ should be roughly $8$ times smaller, and so on. This is in good agreement with our findings (cf. Fig.~\ref{fig:cont_me}). The net effect on $X_{\rm e}$ is about 2.5 times smaller than for $\aEM$ around $z\simeq 1000$ (see Fig.~\ref{fig:compare} for a direct comparison). This suggests that the CMB constraint on $\me$ is weakened by a similar factor. However, adding the rescaling of the Thomson cross section for the computation of the visibility function strongly enhances geometric degeneracies for $\me$, such that the CMB only constraint on $\me$ is $\gtrsim 20$ times weaker than for $\aEM$ (see Sect.~\ref{sec:CMB_me}).

A small difference related to the renormalizations of the photoionization and recombination rates (blue/dashed line) appears. For $\dme>0$, the photoionization rate is increased and the recombination rate is reduced for these contributions [cf. Eq.~\eqref{eq:recfast_scaling}]. Both effects delay recombination (see Fig.~\ref{fig:cont_me}). Thus, around $z\simeq 10^3$ the net effect is slightly larger than for $\aEM$. In contrast to $\aEM$, at late time no net acceleration of recombination occurs. These effects slightly modify the overall redshift dependence of the total $X_{\rm e}$ change, in addition lowering the effect in the freeze-out tail (see Fig.~\ref{fig:compare} for direct comparison). 
At the level $\dme\simeq 1\%$, additional higher order terms become important, allowing one to break the degeneracy between $\aEM$ and $\me$ in joint analyses \citep[see also][]{Planck2015var_alp}.

We note that we ignored the extra $\rho_{\rm b}/\me$ scaling in the Compton cooling term for $T_{\rm e}$, rescaling $\me$ in the atomic quantities only. When varying fundamental constants, dimensionless variables should furthermore be used \citep[e.g.,][]{Rich2015}, so that an analysis of explicit $\me$ variations remains phenomenological.

\vspace{-3mm}
\subsubsection{Comparing \texttt{Recfast++} and \texttt{CosmoRec}}
We close by directly comparing the results for $X_{\rm e}$ obtained with {\tt Recfast++} and {\tt CosmoRec} (Fig.~\ref{fig:compare}). Both codes agree extremely well, departing by $\lesssim 0.1\%$ in $X_{\rm e}$. Tiny differences in the resultant $\Delta X_{\rm e}/X_{\rm e}$ are visible around helium recombination ($z\simeq 1700$), which are related to radiative transfer effects that {\tt CosmoRec} models explicitly. Similarly, around the maximum of the Thomson visibility function ($z\simeq 1100$), small percent-level differences in $\Delta X_{\rm e}/X_{\rm e}$ are present. These differences do not affect the computation of the CMB anisotropies at a significant level and thus our {\tt Recfast++} treatment is sufficient for the analysis presented in Sect.~\ref{sec:constraints}. 
We explicitly confirmed this by comparing the constraints obtained with the two recombination codes for $\aEM$ and $\me$, finding them to agree to high precision. Similarly, for the analysis of future CMB data (e.g. CMB Stage-IV), we deem our treatment with {\tt Recfast++} to suffice in these cases.

\begin{figure}
	\includegraphics[width=\columnwidth]{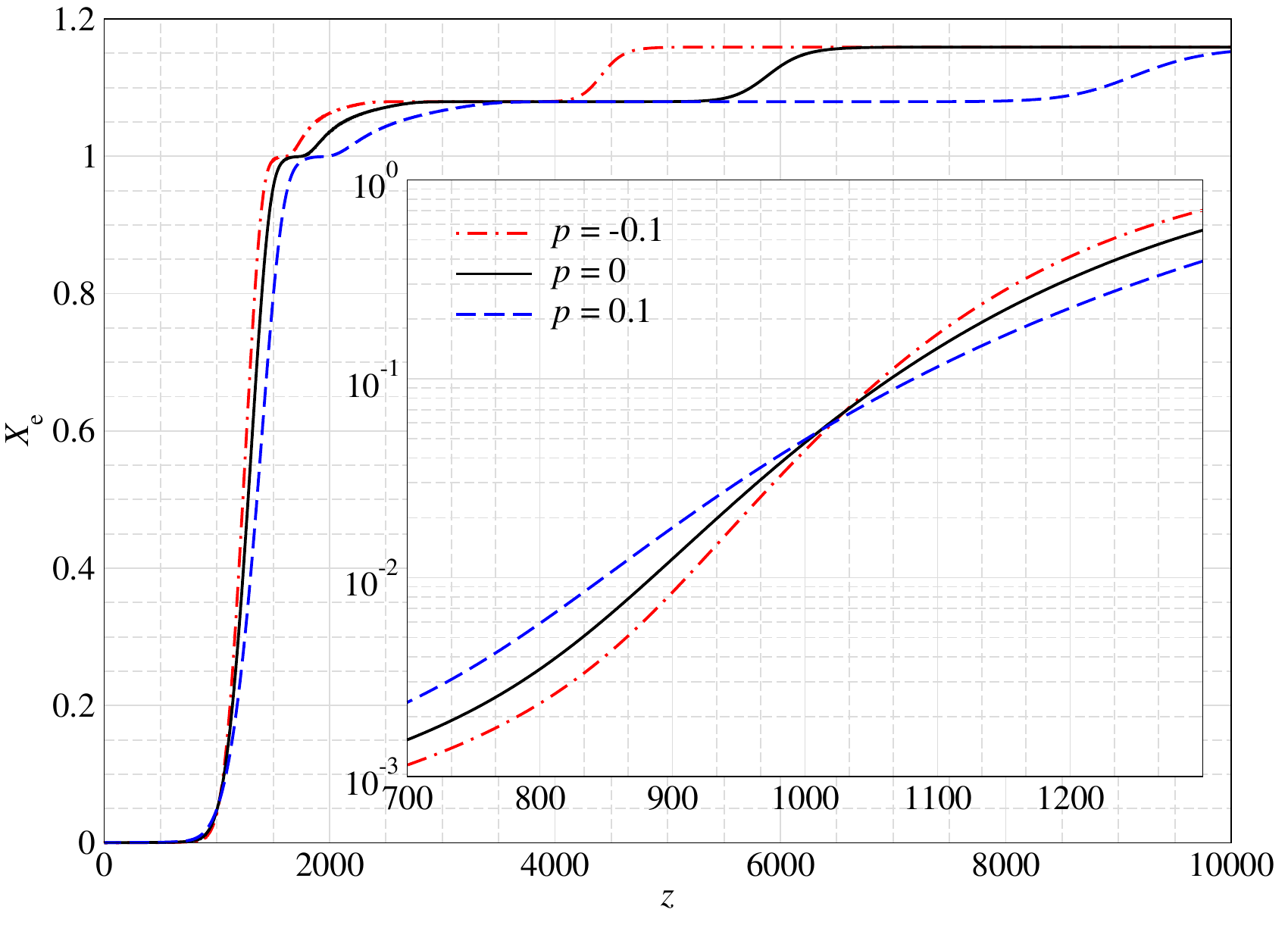}
    \caption{Ionization histories for redshift-dependent variation of the fine-structure constant, $\aEM(z) = \aEM(z_0)\,\left[(1+z)/1100\right]^p$. Here, we set $\aEM(z_0)$ to the standard value, $\aEM(z_0)\approx 1/137$, and only varied $p$. The different phases in the ionization history are stretched/compressed with respect to the standard case, depending on the chosen value for $p\neq 0$.}
    \label{fig:power_law}
\end{figure}

\vspace{-3mm}
\subsection{Adding an explicit redshift dependence to the variations}
\label{sec:time_dep_alpha}
We extend our treatment of variation of fundamental constants by also considering an explicit redshift-dependence of $\aEM$ and $\me$, assuming a phenomenological power-law scaling around pivot redshift\footnote{This choice de-correlates redshift-dependent and constant changes.} $z_0=1100$. This could in principle be caused by the presence of a scalar field and its coupling to the standard particle sector during recombination. For $\aEM$, our model reads
\begin{equation}\label{eq:powerlaw}
  \aEM(z) = \aEM(z_0)\,\left(\frac{1+z}{1100}\right)^p,
\end{equation}
and similarly for $\me$. \changeJ{For $p\ll 1$, we find a logarithmic dependence on redshift, $\aEM(z) \approx \aEM(z_0)\,\left(1+p\ln[(1+z)/1100]\right)$.}
Note that the rescaled value at the pivot redshift is not necessarily $\alpha_{\rm EM}(z_{\rm 0})\equiv\aEMs\simeq 1/137$, but can also be varied. Here, $p$ is a variable index that determines how the ionization history is stretched or compressed around the central redshift. 
We added this \changeJ{new} option to {\tt Recfast++}. Some examples are shown in Fig.~\ref{fig:power_law}. For $p>0$, recombination is accelerated at $z\gtrsim 1000$ with respect to the standard case, while it is delayed at $z\lesssim 1000$. For $p\neq 0$, due to cumulative effects the change in $X_{\rm e}$ does not vanish at the pivot redshift. Also, the modification is very different to that of a constant shift of $\aEM$, predominantly affecting the width of the Thomson visibility function as opposed to the position (see Sect.~\ref{sec:CL_results}). Thus, geometric degeneracies are found to be less important when constraining the value of $p$ using CMB data (Sect.~\ref{sec:constraints}).

\begin{figure}
  \includegraphics[width=\linewidth]{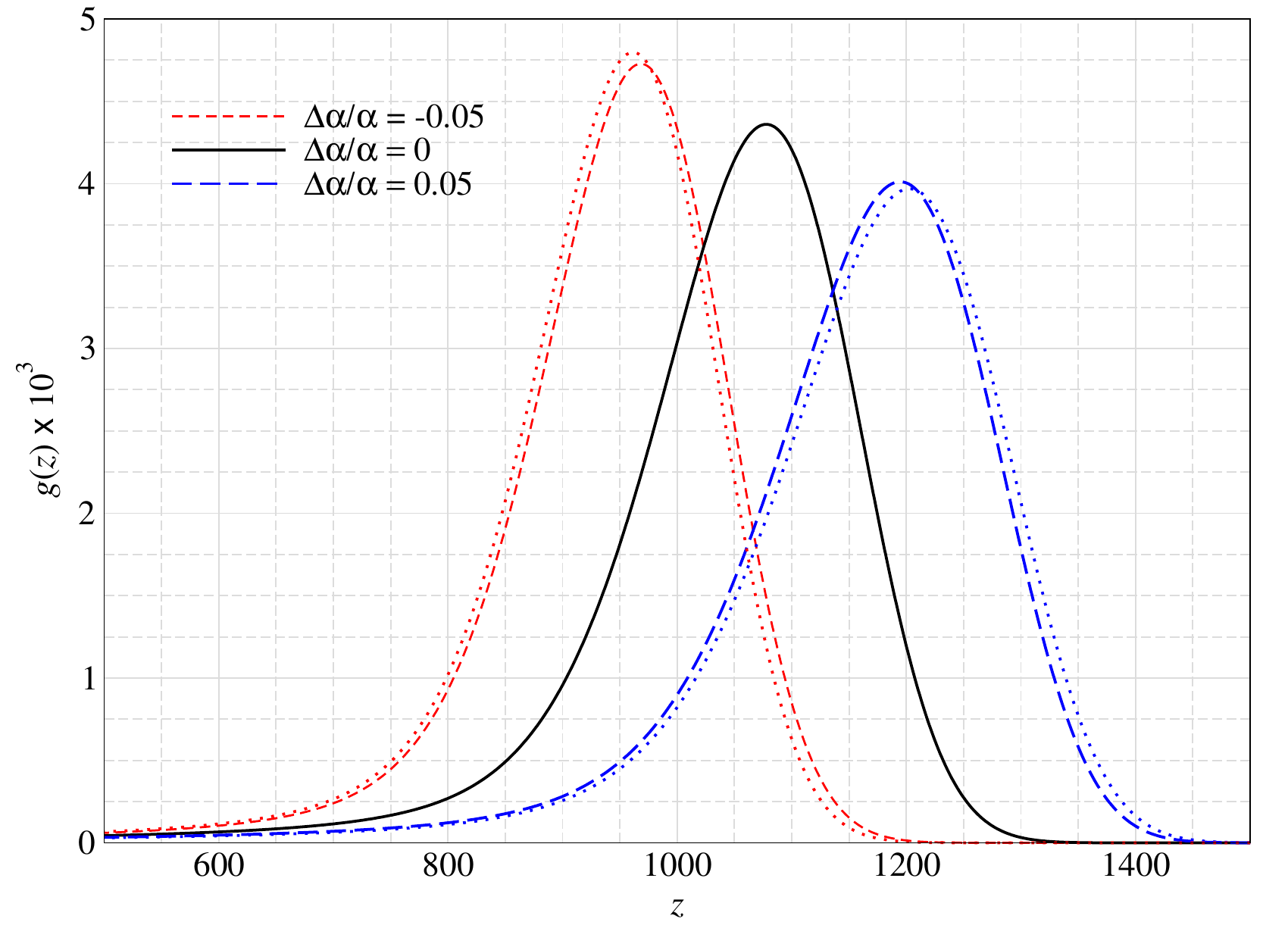}
	  \caption{Visibility functions for a variety of fine structure constant values. A higher value of the fine structure constant leads to a broader visibility function, which simultaneously reduces its height. For illustration, the dotted lines exclude the rescaling of $\sigma_{\rm T}$ within {\tt CAMB}.}
  \label{fig:visi_alpha}
\end{figure}

\vspace{-3mm}
\section{Propagating the effects to the CMB anisotropies}
\label{sec:CL_results}

The temperature and polarization power spectra of the CMB depend on the dynamics of recombination through the ionization history, which defines the Thomson visibility function and last scattering surface \citep[e.g.,][]{Sunyaev1970, Peebles1970, Hu1996anasmall}. Therefore, when varying $\aEM$ and $\me$, this leads to changes in the CMB power spectra. In this section, we show the modifications of the Thomson visibility function for the effects discussed in Section \ref{sec:rec_phys_var}. The modified CMB temperature power spectra are then computed using {\tt CAMB} \citep{CAMB} for the standard cosmology \citep{Planck2015params}. We briefly explain the main effects on the power spectra due to varying fundamental constants. Excluding modifications in the recombination dynamics, the CMB anisotropies still directly depend on the Thomson scattering cross section. We show that the changes from rescaling $\sigma_T$ explicitly within {\tt CAMB} are much smaller than those caused by modifications to the recombination dynamics. Still, they need to be included when deriving CMB constraints on fundamental constants \citep[see also][]{Planck2015var_alp}, in particular when studying variations of $\me$ (see Sect.~\ref{sec:constraints}). We also present the changes of the CMB temperature power spectrum for the redshift-dependent variations of $\aEM$ and $\me$ from Section \ref{sec:time_dep_alpha}. 

\vspace{-3mm}
\subsection{Changes due to constant shifts of $\aEM$ and $\me$}
\label{sec:alpha_cmb}
Using the result for the ionization history computed with the modified version of {\tt Recfast++}, one can calculate the Thomson visibility function, $g(z)$, defined as,
\begin{equation}\label{eq:visi}
  g(z) = \frac{\id \tau}{\id z}\,\exp\left[-\tau(z)\right].
\end{equation}
Here, $\id\tau/\id z$ is the differential Thomson optical depth. The Thomson visibility function can be interpreted as an effective probability distribution for a photon being last-scattered around redshift $z$. It is normalized such that $\int g(z) \id z = 1$. From the changes in $X_{\rm e}$ described above, we expect that for $\dalpha>0$ the maximum of the visibility function shifts toward higher redshifts. In Fig.~\ref{fig:visi_alpha}, the visibility function is shown for constant $\dalpha = \{-0.05,0,0.05\}$. Indeed, the visibility function maximum moves to $z^{\rm max}\approx 1200$ for $\dalpha=0.05$. The relative width, $\Delta z^{\rm \;FWHM}/z^{\rm \,max}$, of the visibility function is roughly conserved.

\vspace{-3mm}
\subsubsection{Effects on the CMB anisotropies due to variations of $\aEM$}
\label{sec:CMB_alpha}
\begin{figure}
	\includegraphics[width=\columnwidth]{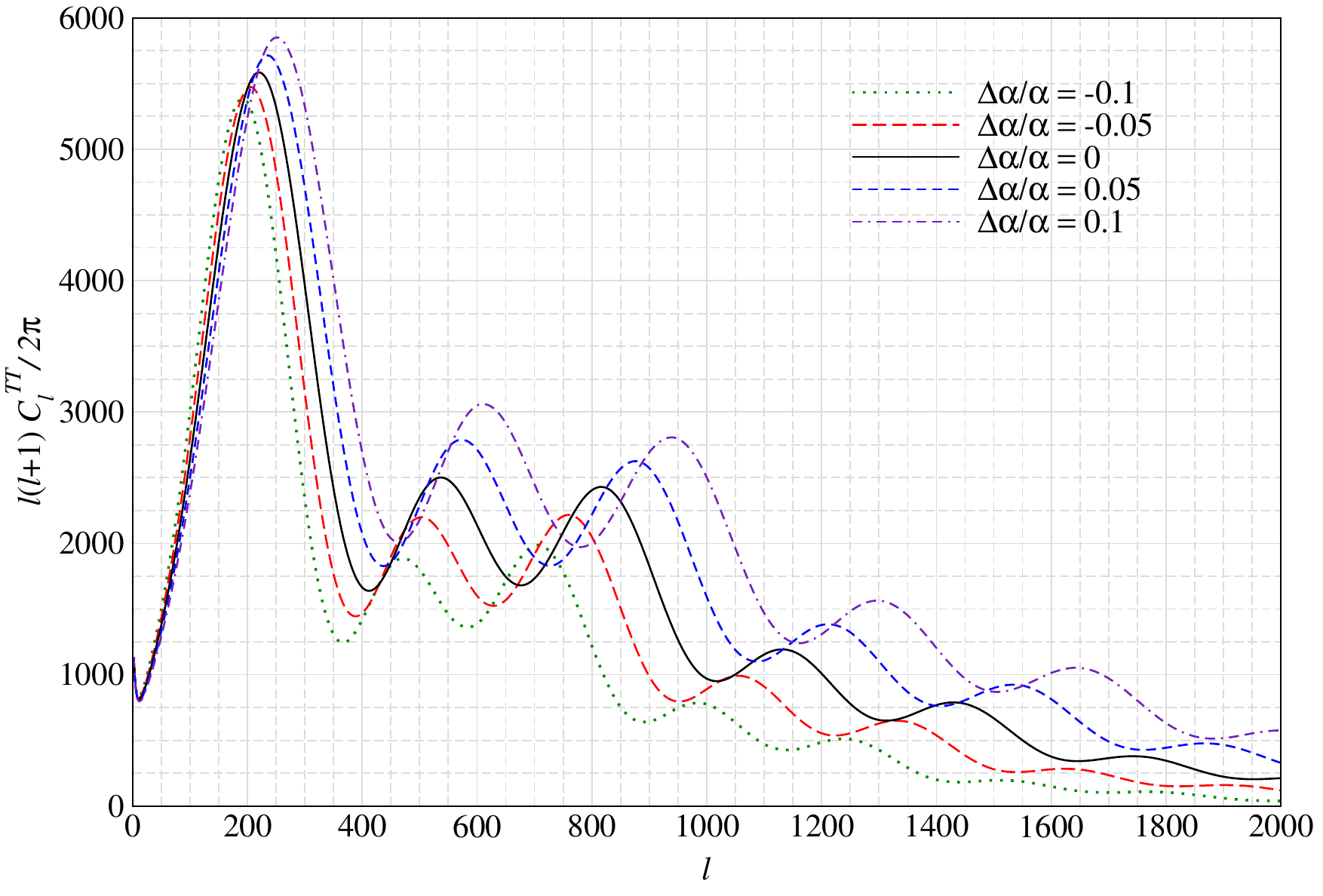}
    \caption{The CMB temperature power spectra for different values of $\aEM$. This shows that as the fine structure constant increases, the anisotropies shift toward smaller scales and higher amplitudes.}
    \label{fig:alpha_cl}
\end{figure}

We illustrate the modifications to the CMB power spectrum for constant changes of $\aEM$ in Fig.~\ref{fig:alpha_cl}. We focus on the CMB temperature power spectra, as the effects on the polarization power spectra are qualitatively similar. Two main effects are visible. Firstly, the peaks of the power spectrum are shifted to smaller scales (larger $\ell$) when $\dalpha >0$. This happens because earlier recombination moves the last scattering surface towards higher redshifts, which decreases the sound horizon and increases the angular diameter distance to recombination \citep{Kaplinghat1999, Battye2001}. 
Secondly, for $\dalpha>0$, the peak amplitudes are enhanced. This is mainly because earlier recombination suppresses the effect of photon diffusion damping on the anisotropies \citep{Kaplinghat1999, Battye2001}.
For small $\dalpha$, we also illustrate the relative change of the temperature power spectrum in Fig.~\ref{fig:talpha}. The effect on the peak positions is more noticeable than the small overall tilt caused by changes related to diffusion damping.

\begin{figure}
	\includegraphics[width=\columnwidth]{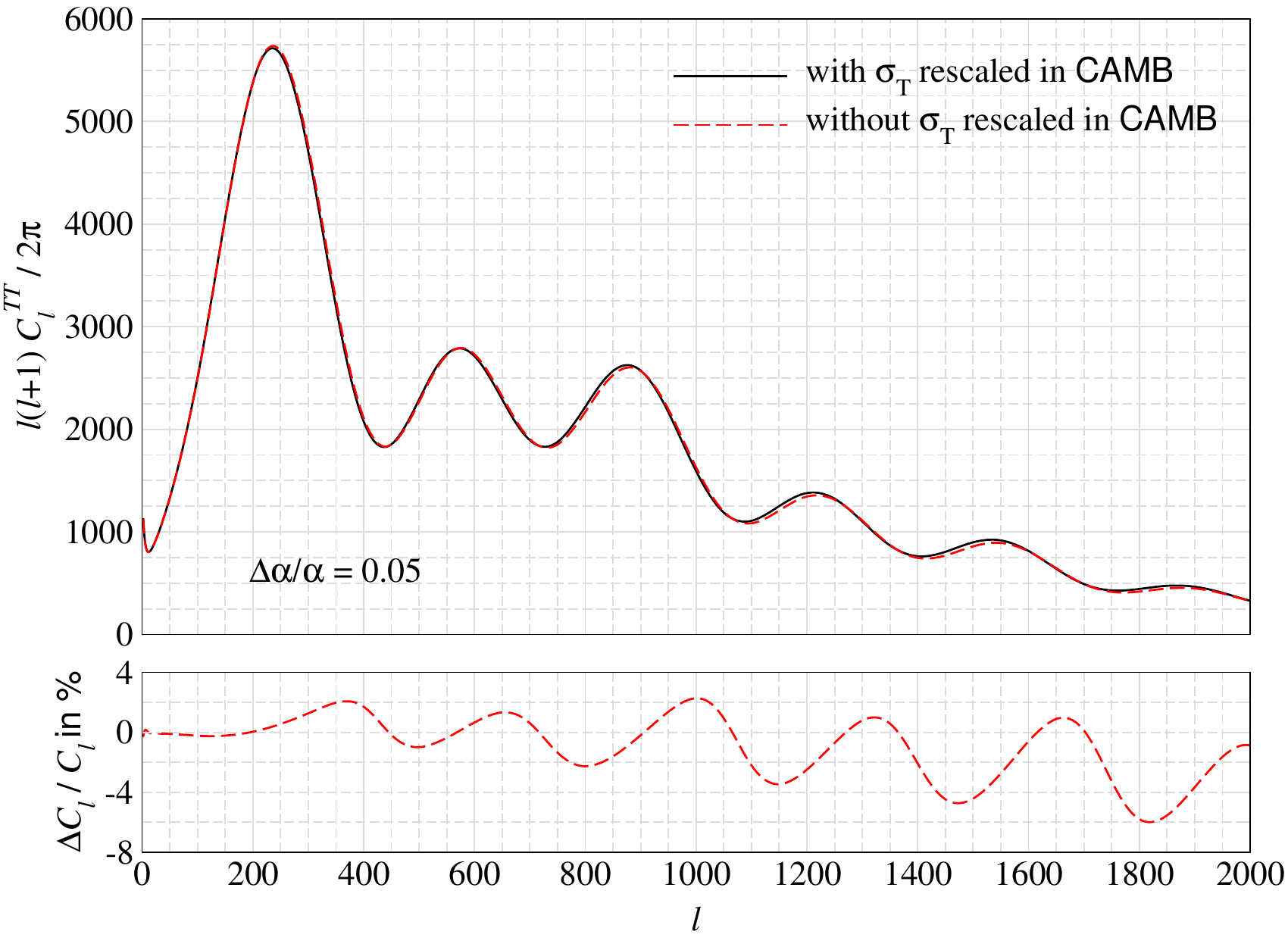}
	\caption{The CMB temperature power spectrum for $\Delta \alpha/\alpha=0.05$ computed with and without explicit $\sigma_{\rm T}$ rescaling within {\tt CAMB}. The upper panel shows the temperature power spectra and the lower illustrates the corresponding relative difference with respect to the full case.}
	\label{fig:thom_all}
\end{figure}

\vspace{-3mm}
\subsubsection{Separate effect related to $\sigT$}
\label{sec:SigT}
The Thomson scattering cross section, $\sigT$, enters the problem in two ways. Firstly, it directly affects the recombination dynamics and thermal coupling between photons and baryons, as explained above (Sect.~\ref{sec:effect_Xe}). These changes are taken into account when computing the recombination history, but turn out to be minor (Fig.~\ref{fig:cont} and \ref{fig:cont_me}) and can in principle be neglected. 
Secondly, $\sigT$ also directly appears in the definition of the Thomson visibility function, $g(z)$, which is computed inside {\tt CAMB} and has to be modified separately \citep[see][]{Kaplinghat1999, Planck2015var_alp}. The comparably small effect on $g(z)$ is illustrated in Fig.~\ref{fig:visi_alpha} for $\dalpha=\pm 0.05$, where the dotted lines exclude the rescaling of $\sigma_{\rm T}$ within {\tt CAMB}.
The corresponding changes to the $TT$ power spectrum for $\dalpha=0.05$ are shown in Fig.~\ref{fig:thom_all}. In the considered $\ell$ range, the maximal relative difference is \changeJ{$|\Delta C_{\ell}/C_{\ell}| \simeq 6\%$}, which is more than one order of magnitude smaller than the effects due to direct changes in $X_{\rm e}$ discussed above. However, in particular when studying variations of $\me$, this effect has to be included as otherwise the errors are strongly underestimated (see Sect.~\ref{sec:constraints}).

We included the effect of $\sigT$ rescaling for the computation of the visibility in two independent ways. First, we consistently implemented these changes into {\tt CAMB} by adding a rescaling function that targets the {\tt akthom} components and Compton cooling terms within {\tt modules.f90} and {\tt reionization.f90}. Second, we simply redefined the free electron fraction, $X_{\rm e}$ returned by the recombination code to {\tt CAMB} as $X_{\rm e}^*=(\aEM/\aEMs)^2(\me/\mes)^{-2}\,X_{\rm e}$. The two approaches gave extremely similar results for the power spectra and also final parameter constraints. The only real difference is that in the first approach, the corrections to the reionization history are included more consistently, albeit not being modeled in a physical manner. For example, for $\dalpha>0$, the reionization redshift reduces for fixed value of $\tau$. In the second approach, the reionization history is not affected but the correction is minor. For our analysis, we used the explicit rescaling in {\tt CAMB} including all terms.

\vspace{-4mm}
\subsubsection{Effects on the CMB anisotropies due to variations of $\me$}
\label{sec:CMB_me}
We now briefly mention the changes caused by variation of $\me$. As discussed above, for the free electron fraction the net changes are very similar to those for $\aEM$. Thus, one expects both changes in the visibility and CMB power spectra to be similar, albeit at a lower amplitude when $\dme \simeq \dalpha\ll 1$. Indeed, we find the changes in the visibility function around its maximum to mimic those shown in Fig.~\ref{fig:visi_alpha} for variations of $\aEM$ when setting $\dme \simeq (2-3)\times \dalpha$. This is expected when comparing the main effect on $X_{\rm e}$ around redshift $z\simeq 10^3$ for $\aEM$ and $\me$ (Fig.~\ref{fig:cont} and Fig.~\ref{fig:cont_me}), and suggests that the $\me$-related changes in the CMB power spectra are also weakened by a similar factor.
\begin{figure}
  \includegraphics[width=\linewidth]{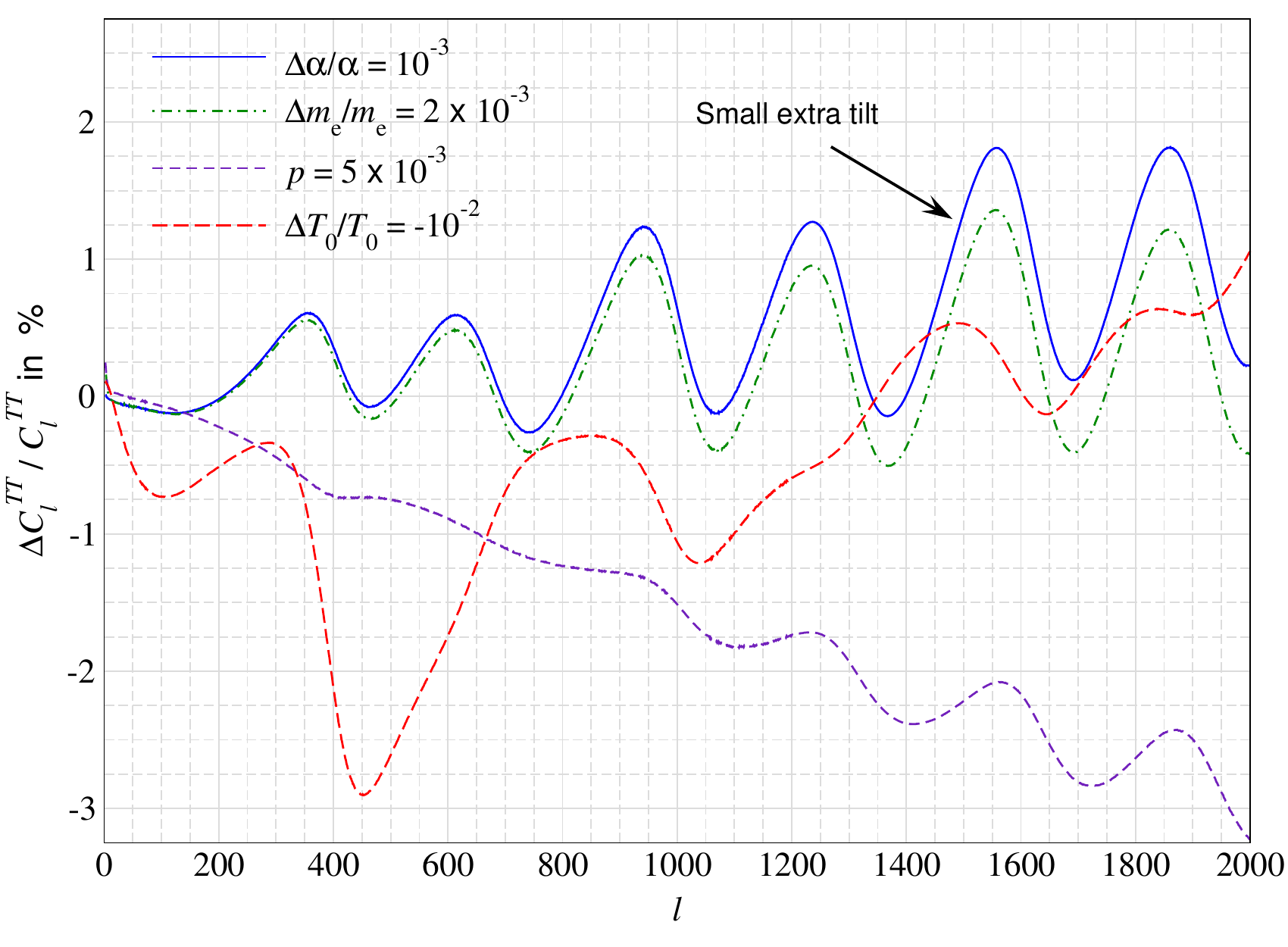}
  \caption{Comparison of the CMB $TT$ power spectrum deviations when varying $\aEM$, $\me$, $T_0$ and $p$. We chose $\dalpha=10^{-3}$, $\dme=2\times 10^{-3}$, $\Delta T/T=-10^{-2}$ and \changeJ{$p=5\times 10^{-3}$ (simultaneously for $\aEM$ and $\me$)} to obtain effects at a similar level. Notice the small extra tilt when comparing the case for $\aEM$ with $\me$, which helps when constraining $\aEM$.}
  \label{fig:talpha}
\end{figure}
This is explicitly illustrated in Fig.~\ref{fig:talpha}, which shows that aside from a small overall tilt the changes in the CMB $TT$ power spectra, $\Delta C_{\ell}/C_{\ell}(\dalpha)$ and $\Delta C_{\ell}/C_{\ell} (\dme)$, become almost indistinguishable when using $\dme\approx (2-3)\,\dalpha$.
This presents a quasi-degeneracy between the two parameters and also suggests that naively the analysis for $\dalpha$ could be sufficient to estimate the errors for a corresponding analysis of $\me$. 
However, when constraining $\me$, enhanced geometric degeneracies (because of $\sigT$) push the error to the percent level. In this case, higher order terms become important and the degeneracy is again broken. When also adding information from BAO, the error on $\me$ is strongly reduced. In this regime, we indeed recover the simple scaling of the errors, $\sigma(\dme)\simeq 3\sigma(\dalpha)$ [see Table~\ref{table:param_me}].

\vspace{-4mm}
\subsubsection{Degeneracies between $\aEM$ and $T_0$}
\label{sec:T0_degen}
Our previous discussion showed that a variation of $\aEM$ and $\me$ directly affect the recombination redshift. The main effect can be captured by rescaling the Boltzmann factors using $T_{\rm eff}$. This suggests that a change in the CMB monopole temperature, $T_0$, could have a very similar effect. However, there is one crucial difference: $T_{\rm 0}$ also affects the matter-radiation equality epoch. This modifies the integrated Sachs Wolfe (ISW) effect, which is most noticeable at low and intermediate $\ell$ and in principle should help one to break the degeneracy between changes of $\aEM$ and $T_{\rm 0}$. 
The relative change to the $TT$ power spectrum is illustrated in Fig.~\ref{fig:talpha}, where we use $\Delta T_0/T_0\simeq -0.01$ to make the modifications comparable in amplitude. One can clearly see the enhanced effect at large angular scales due to the early and late ISW \citep[see Fig.~3 in][for illustration of these contributions]{Challinor2009}.

\begin{figure}
  \includegraphics[width=\linewidth]{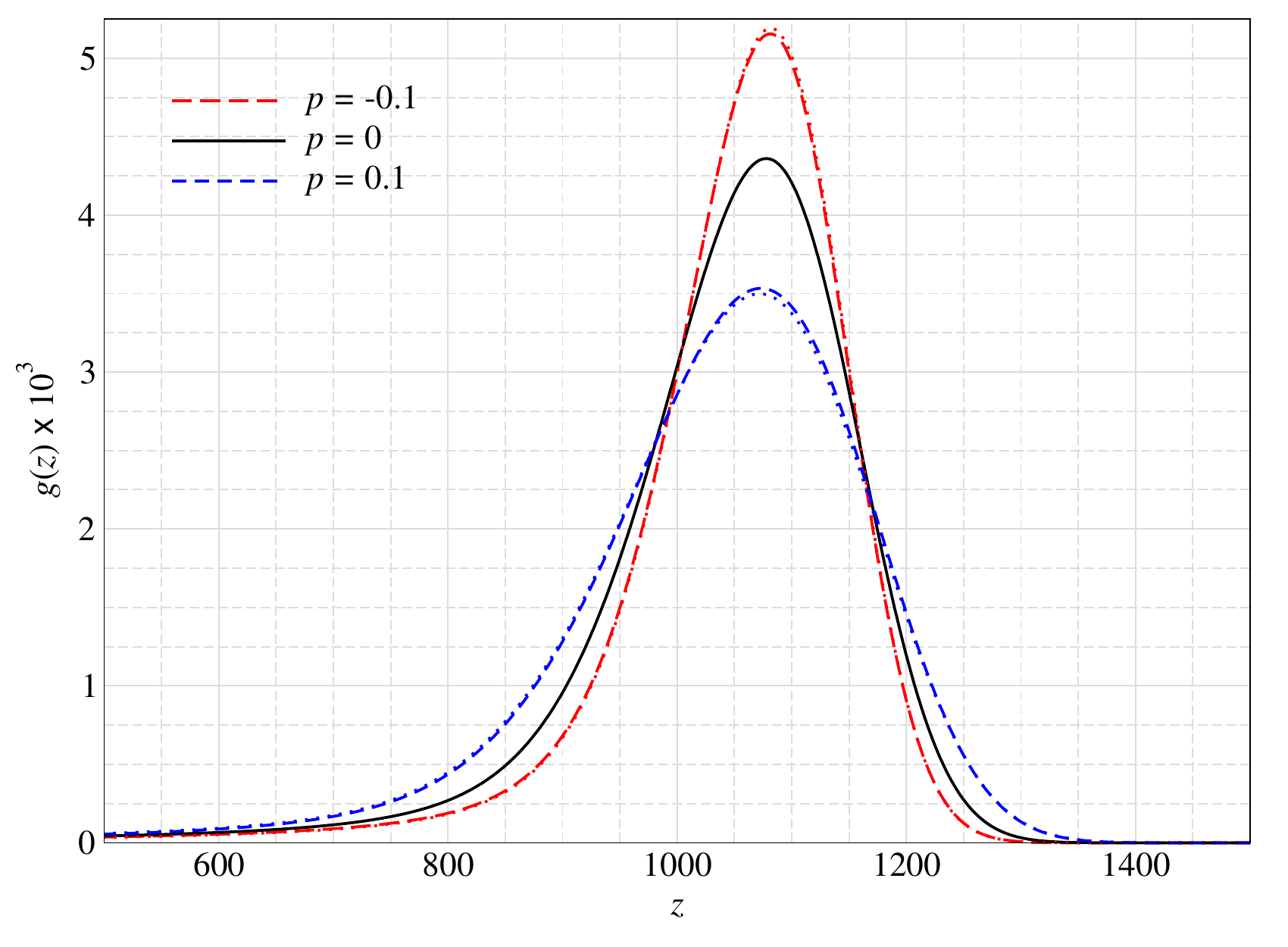}
  \caption{The visibility function $g(z)$ of the CMB for various cases of redshift dependent $\aEM$, parametrized as in Eq.~\eqref{eq:powerlaw}. For $p<0$, the recombination era is confined to a narrower redshift range as shown in Fig.~\ref{fig:power_law}, an effect that narrows the visibility function. Again, the dotted lines exclude the rescaling of $\sigma_{\rm T}$ within {\tt CAMB}, which has a small overall impact here.}
\label{fig:visi_power}
\end{figure}
\begin{figure}
	\includegraphics[width=\columnwidth]{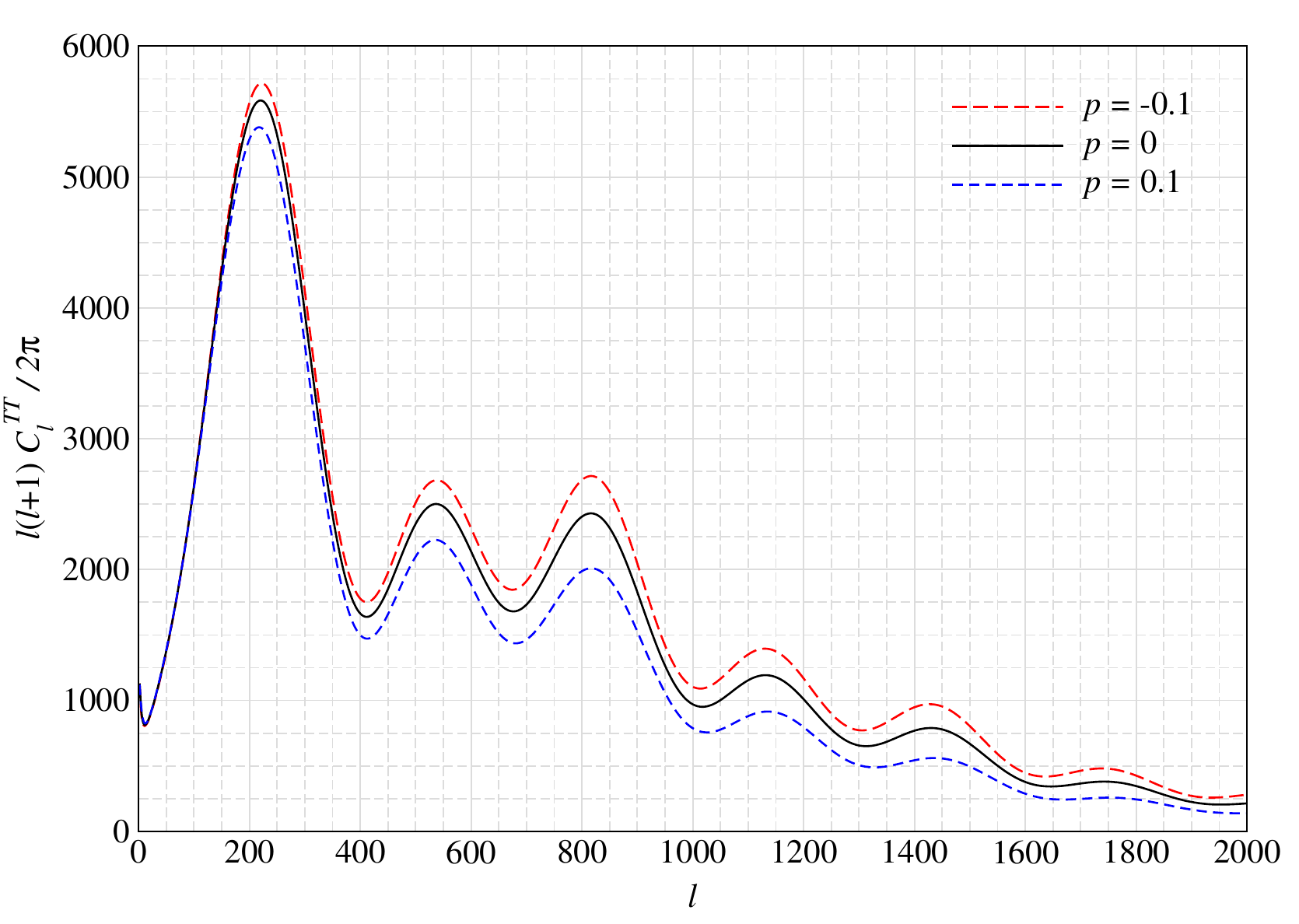}
    \caption{The CMB power spectrum for various values of the redshift power law index $p$ for $\aEM$. Positive values of the $p$ index lead to a suppression of the CMB peaks due to broadening of the recombination epoch.}
    \label{fig:power_cl}
\end{figure}

\begin{table*}
  \centering
\begin{tabular} { l  c c c c}
\hline\hline
 Parameter &  {\it Planck} 2015 & + varying $\aEM/\aEMs$ & + varying $p$ & + varying $\aEM/\aEMs$ and $p$\\
\hline
$\Omega_{\rm b} h^2   $ & $0.02224\pm 0.00016 $& $0.02225\pm 0.00016$ & $0.02226\pm 0.00018$ & $0.02223\pm 0.00019      $\\

$\Omega_{\rm c} h^2   $ & $0.1193\pm 0.0014$& $0.1191\pm 0.0018$ & $0.1194\pm 0.0014$ & $0.1193\pm 0.0020        $\\

$100\theta_{\rm MC} $ & $1.0408\pm 0.0003$ & $1.0398\pm 0.0035$ & $1.0408\pm 0.0003$ & $1.0406\pm 0.0051         $\\

{\boldmath$\tau           $} & $0.062\pm 0.014$& $0.063\pm 0.014$ & $0.062\pm 0.014$ & $0.063\pm 0.015           $\\

${\rm{ln}}(10^{10} A_{\rm s})$ & $3.057\pm 0.025$ & $3.060\pm 0.027$ & $3.058\pm 0.026$ & $3.059\pm 0.027           $\\

$n_{\rm s}            $ & $0.9649\pm 0.0047$ & $0.9668\pm 0.0081$ & $0.9663\pm 0.0060$ & $0.9666\pm 0.0081        $\\

\hline
$\aEM/\aEMs$ & -- & $0.9993\pm 0.0025$ & -- & $0.9998\pm 0.0036        $\\

$p$ & -- & -- & $0.0008\pm 0.0025$ &  $0.0007 \pm 0.0036$\\

\hline
$H_0 \,[{\rm km \, s^{-1}\,\Mpc^{-1}}]    $ & $67.5\pm 0.6$& $67.2\pm 1.0$ & $67.5\pm 0.6$ & $67.3\pm 1.4             $\\
\hline\hline
\end{tabular}
\caption{Constraints on the standard $\Lambda$CDM parameters and the fundamental constant parameters $\aEM/\aEMs$ and $p$ for different combinations of parameters. The standard {\it Planck} runs include the $TTTEEE$ likelihood along with the low $\ell$ polarization and CMB lensing likelihoods and the errors are the $68\%$ limits.}
\label{table:param}
\end{table*}

\vspace{-3mm}
\subsection{Changes due to redshift-dependent variations}
\label{sec:power_cmb}
We now consider redshift-dependent variations to $\aEM$ and $\me$, using the parametrization given by Eq.~\eqref{eq:powerlaw}. We assume the standard values for $\aEM$ and $\me$ at $z_0=1100$. In Fig.~\ref{fig:visi_power}, we illustrate the effect on the Thomson visibility function for $\aEM$. Using $p<0$, the visibility function narrows such that the effective width, $\Delta z^{\rm \,FWHM}/z^{\rm \,max}$, reduces. However, the changes in the position of the maximum value of $g(z)$ are negligible. This suggests that the changes in the positions of the peaks in the CMB power spectra are minor, while the {\it blurring} related to the finite thickness of the last scattering surface is reduced\footnote{An explanation of this damping effect can be found in \cite{Mukhanov2004}.}. 
The separate effect of rescaling $\sigT$ inside {\tt CAMB} is also illustrated in Fig.~\ref{fig:visi_power}, an effect that we find to have a negligible impact when constraining the value of $p$ alone. 
Similar comments apply for changes to $\me$. 

We show the changes in the CMB temperature power spectra due to redshift-dependent variations of $\aEM$ in Fig.~\ref{fig:power_cl}. When we choose $p<0$, the CMB peaks are amplified. This is expected from the reduced width, $\Delta z^{\rm \,FWHM}/z^{\rm \,max}$, of the visibility function in Fig.~\ref{fig:visi_power}. Similarly, for $p>0$, a larger damping effect due to blurring is found. The relative change of $C_{\ell}$ for $p=5\times10^{-3}$ is shown in Fig.~\ref{fig:talpha}. The smoothness of the titled curve indicates that blurring of anisotropies is indeed the dominant effect. Again, similar effects are found for changes to $\me$.

\begin{figure*}
  \includegraphics[width=0.8\textwidth]{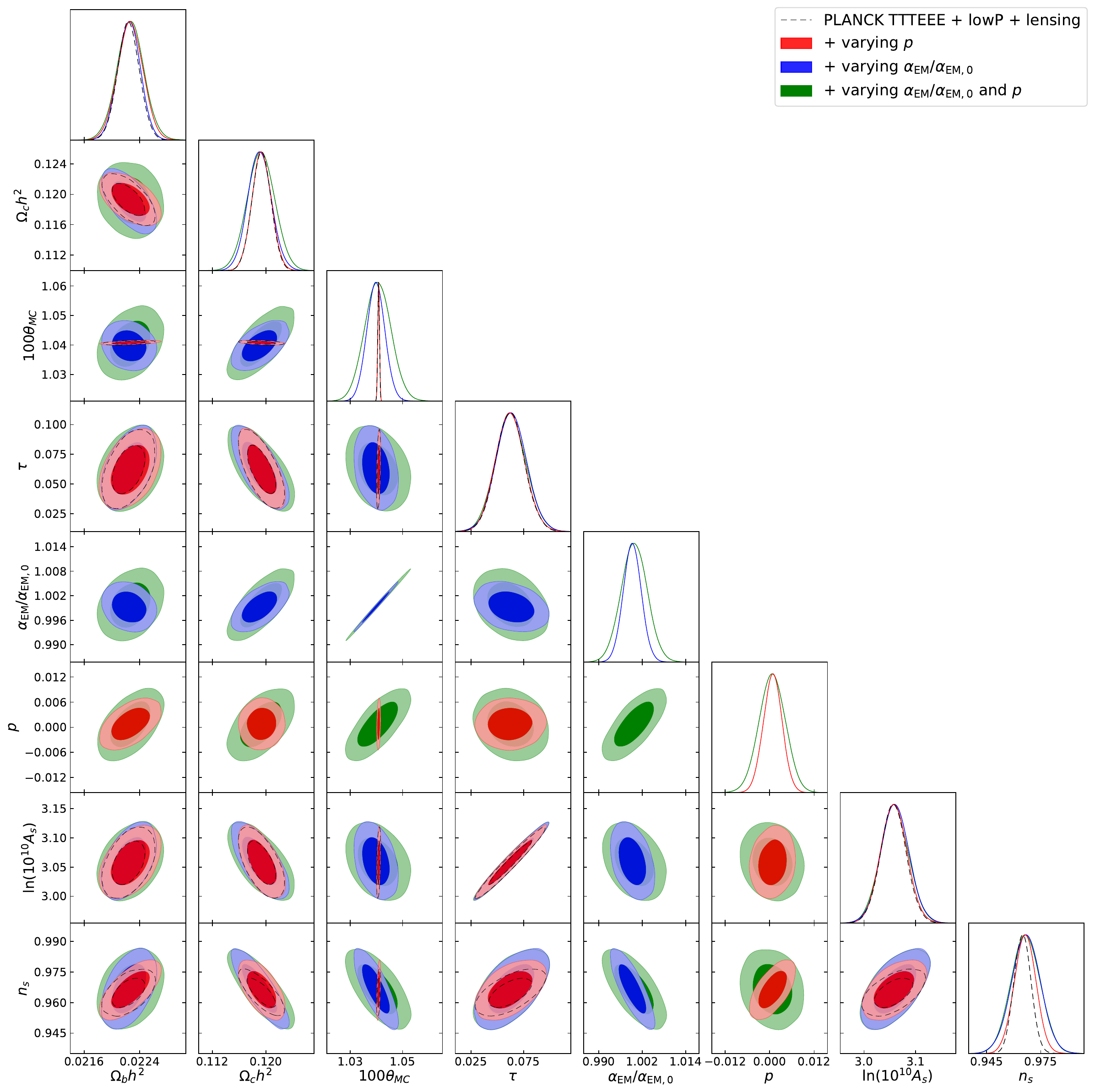}
  \caption{The full contours with the $TTTEEE$ + lowP + lensing likelihoods from {\it Planck} 2015. The standard \LCDM~ run {\it (black, dotted)} is shown alongside the added variations in $\aEM$ {\it (blue)}, $p$ {\it (red)} and the combination of the two parameters {\it (green)}.}
 \label{fig:contour} 
\end{figure*}

\vspace{-4mm}
\section{Constraints using PLANCK data}
\label{sec:constraints}
We now constrain the variations of $\aEM$, $\me$ and $p$ discussed in Sec.~\ref{sec:CL_results} using {\tt CosmoMC} \citep{COSMOMC} with the {\it Planck} 2015 data\footnote{When quoting {\it Planck} 2015 data we usually refer to the likelihood {\it Planck} 2015 $TTTEEE$+lowP+lensing. For {\it Planck 2013}, we imply {\it Planck}+WP+lensing as baseline.}. We sample over the acoustic angular scale, $\theta_{\rm MC}$. Although for this specific analysis, $H_0$ is expected to de-correlate quicker, we did not encounter any problems. We find that the constraints derived for $\aEM$ and $\me$ are consistent with those for {\it Planck} 2013 data \citep{Planck2015var_alp}, albeit here with slightly improved errors. We also show the new constraints for our redshift dependent model of $\aEM$. 
Our marginalized constraints are summarized in Tables \ref{table:param} and \ref{table:param_me}. For comparison, the standard 6 \LCDM~parameter run for the {\it Planck} data is also given. For each run, we show the derived $H_0$ parameter. The 2D parameter contours are shown in Fig.~\ref{fig:contour}.

\begin{table*}
  \centering
\begin{tabular} { l  c c c c}
\hline\hline
 Parameter 
& {\it Planck} 2015 
& {\it Planck} 2015 + BAO
& {\it Planck} 2015  
& {\it Planck} 2015 + BAO
\\
& + $\me/\mes$ 
& + $\me/\mes$ 
& + $\aEM/\aEMs$ and $\me/\mes$
& + $\aEM/\aEMs$ and $\me/\mes$
\\
\hline

$\Omega_{\rm b} h^2$ & $0.0213_{-\,0.0017}^{+\,0.0011}$ & $0.02233\pm \
0.00018$ & $0.0214_{-\,0.0017}^{+\,0.00099}$ & $0.02238\pm 0.00020$ \\[0.5mm] 
$\Omega_{\rm c} h^2$ & $0.1146_{-\,0.0087}^{+\,0.0059}$ & $0.1202\pm \
0.0022$ & $0.1144_{-\,0.0090}^{+\,0.0057}$ & $0.1200\pm 0.0023$ \\[0.5mm]  
$\theta_{\rm MC}$ & $1.012_{-\,0.051}^{+\,0.036}$ & $1.0435\pm \
0.0052$ & $1.011_{-\,0.054}^{+\,0.034}$ & $1.0431\pm 0.0053$ \\[0.5mm]  
$\tau$ & $0.057\pm 0.015$ & $0.080\pm 0.017$ & $0.058\pm 0.015$ & \
$0.082\pm 0.019$ \\[0.5mm]  
$\ln(10^{10} A_s)$ & $3.044\pm 0.029$ & $3.095\pm 0.033$ & $3.048\pm \
0.031$ & $3.100\pm 0.037$ \\[0.5mm] 
$n_s$ & $0.9639\pm 0.0048$ & $0.9647\pm 0.0046$ & $0.9663\pm 0.0078$ \
& $0.9678\pm 0.0085$ \\[0.5mm]  
\hline
$\alpha_{\rm EM}/\alpha{\rm EM,0}$ & --  & --  & $0.9990\pm 0.0025$ & \
$0.9989\pm 0.0026$ \\[0.5mm]  
$m_{\rm e}/m_{\rm e,0}$ & $0.961_{-\,0.072}^{+\,0.046}$ & $1.0039\pm \
0.0074$ & $0.962_{-\,0.074}^{+\,0.044}$ & $1.0056\pm 0.0080$ \\[0.5mm] 
\hline
$H_0\,{\rm [km\,s^{-1}\,Mpc^{-1}]}$ & $60_{-\,16}^{+\,8}$ & $68.1\
\pm 1.3$ & $60_{-\,16}^{+\,7}$ & $68.1\pm 1.3$ \\ 
\hline\hline
\end{tabular}
\caption{Constraints on the standard $\Lambda$CDM parameters and the fundamental constant parameters $\aEM/\aEMs$ and $\me/\mes$ for different combinations of parameters. The standard {\it Planck} runs include the $TTTEEE$ likelihood along with the low $\ell$ polarization and CMB lensing likelihoods and the errors are the $68\%$ limits. We also show the results when adding BAO data. }
\label{table:param_me}
\end{table*}


\vspace{-3mm}
\subsection{Constraining $\aEM$ and $\me$}
\label{sec:alpha_me}
When varying $\aEM$, assuming constant $\dalpha$, along with the 6 standard cosmological parameters, we find the marginalized parameter values in the second column of Table~\ref{table:param}. These show that $\aEM/\aEMs$ is equal to unity well within the $68\%$ limit. The errors on $\theta_{\rm MC}$ increases by about one order of magnitude, due to the added uncertainty in the distance to the last scattering surface. We also find a slight increase in the errors of the scalar spectral index, $n_{\rm s}$, which interacts with the modifications to the photon diffusion damping scale caused by $\aEM$. Similarly, the error of the cold dark matter density, $\Omega_{\rm c} h^2$, increases slightly, due to geometric degeneracies. The other parameters (i.e., $\tau$ and $A_{\rm s}$) are largely unaltered by the addition of $\aEM$ as a parameter (see Fig.~\ref{fig:contour}). This highlights the stability and consistency of the data with respect to non-standard extensions of the cosmological model. 

Although the contributions from $\sigT$ appear to have a negligible effect on the $C_\ell$ (see Fig.~\ref{fig:thom_all}), we find that the inclusion of this effect improves the errors on $\aEM$ by $\simeq 30\%$. The small effect on the power spectra hinders some of the degeneracy between $\theta_{\rm MC}$ and $\aEM/\aEMs$, as pointed out in previous analyses \citep[e.g.][]{Planck2015var_alp}. For example, the marginalized value of $\aEM/\aEMs$ changes from $0.9988 \pm 0.0033$ to $0.9993 \pm 0.0025$ when including $\sigT$ rescaling within {\tt CAMB}\footnote{The former result was also presented in the initial versions of \citet{CORE2016}, but now agrees with latter.}. Future CMB missions have the potential to improve this constraint by another factor of $\simeq 4-5$ \citep{CORE2016}.

For constant changes to $\me$, we obtain the results given in the first two columns of Table~\ref{table:param_me}. Using 2015 CMB data alone, we find $\me/\mes=0.961^{+0.046}_{-0.072}$ and $H_0=60^{+\,8}_{-16}\,{\rm km \, s^{-1}\,\Mpc^{-1}}$, which is consistent with the corresponding result ({\it Planck}+WP+lensing), $\me/\mes=0.969 \pm 0.055$ and $H_0=(62\pm 10)\,{\rm km \, s^{-1}\,\Mpc^{-1}}$, given in \citet{Planck2015var_alp}. Following the {\it Planck} 2013 analysis, we used a flat prior $H_0=[40, 100]\,{\rm km \, s^{-1}\,\Mpc^{-1}}$. We note that at the lower end of this range this leads to a slight truncation of the posterior distribution for $H_0$. Also, our errors for the CMB-only analysis remain asymmetric, even when repeating the {\it Planck} 2013 run, for which we find $\me/\mes=0.964^{+0.054}_{-0.068}$ and $H_0=61^{+9.5}_{-15}\,{\rm km \, s^{-1}\,\Mpc^{-1}}$. The remaining difference between the errors is not negligible and should be investigated further. We already confirmed that it is not related to the slightly different scalings for $\alpha_{\rm rec}$ and $\beta_{\rm phot}$ used in \citet{Planck2015var_alp}, but leave a more detailed study to future work.

For varying $\me$, the values of $H_0$ and $\me$ are both biased low when only using CMB data \citep[see also][]{Planck2015var_alp}. Interestingly, this bias is removed when neglecting the effect of $\sigT$ on the Thomson visibility, for which we find $H_0=(67.0\pm 1.6)\,{\rm km \, s^{-1}\,\Mpc^{-1}}$ and $\me/\mes= 0.9970\pm 0.0098$. This treatment also significantly decreases the errors due to reduced geometric degeneracies, which highlights the importance of $\sigT$ for the computation of constraints on variations of $\me$. At the level of $\dme\simeq 1\%$, also non-linear corrections become noticeable.

Given the large geometric degeneracy, we also ran the constraint for $\me$ when adding BAO data (see Table~\ref{table:param_me}). In this case, we obtained\footnote{\changeJ{Adding lensing did not affect the constraint at a very significant level.}} $\me/\mes= 1.0039\pm 0.0074$, which, albeit improved error, is consistent with the result $\me/\mes= 1.004 \pm 0.011$ given in \citet{Planck2015var_alp}. The value of $H_0$ returns to the standard CMB value when adding BAO data. For this combination of datasets, we also find that the error on $\me$ is $\simeq 3$ times larger than the corresponding error on $\aEM$, as naively expected from the similarity of the changes in the $TT$ power spectrum (Fig.~\ref{fig:talpha}).

\begin{figure}
	\includegraphics[width=\columnwidth]{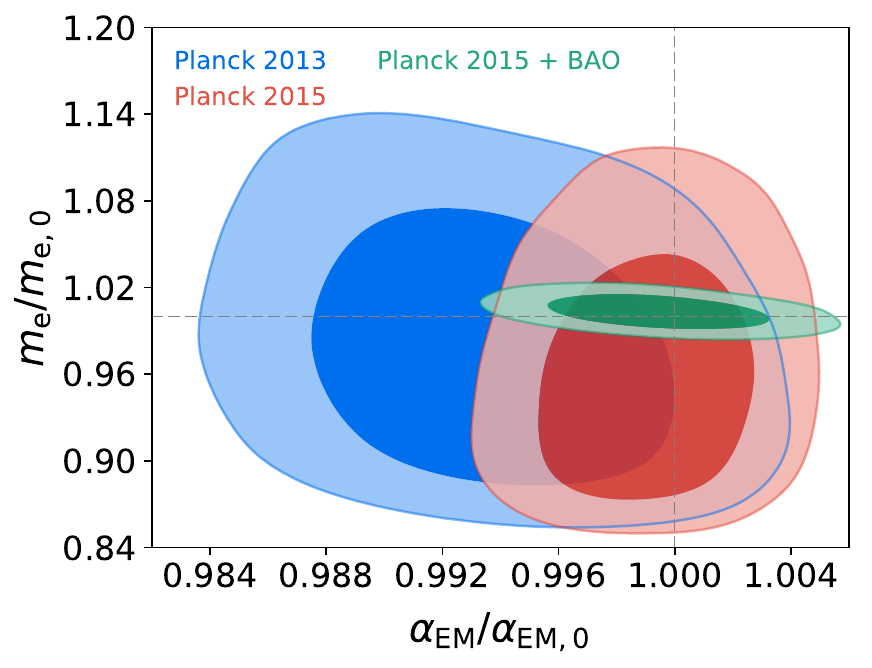}
    \caption{The joint likelihood contour for $\me$ and $\aEM$ for {\it Planck} 2013+WP+lensing and {\it Planck} 2015 $TTTEEE$+lowP+lensing data. The {\it Planck} 2015 contours are narrowed in the $\aEM$ direction due to improved polarization information over {\it Planck} 2013. Adding BAO data to {\it Planck} 2015 further improves the constraint in particular on $\me$. The dashed lines indicate $\dme=\dalpha=0$ for reference.}
    \label{fig:malph}
\end{figure}

\vspace{-3mm}
\subsubsection{Simultaneously constraining $\aEM$ and $\me$}
\label{sec:alpha_me_sim}
We finish our analysis of this section by simultaneously varying $\aEM$ and $\me$ (see
Table~\ref{table:param_me} for our constraints). The responses in the CMB power spectra are quite similar for $\dalpha\simeq \dme\simeq 10^{-3}$ (see Fig.~\ref{fig:talpha}), suggesting a significant degeneracy between $\aEM$ and $\me$. However, combined CMB-only constraints are obtained when non-linear corrections in particular for $\me$ become noticeable, so that both parameters can be simultaneously constrained. 

The strong degeneracies between $\aEM$ and $\me$ are substantially reduced when going from {\it WMAP} to {\it Planck} 2013, as already described in \citet{Planck2015var_alp}. In Fig.~\ref{fig:malph}, we show our contours for {\it Planck} 2013 and 2015 data. The non-Gaussian shapes of the contours are reminiscent of the non-linear terms mentioned above. We find $\aEM/\aEMs=0.9990\pm 0.0025$ and $\me/\mes=0.962^{+0.044}_{-0.074}$ for {\it Planck} 2015. This improves over our constraint for {\it Planck} 2013, $\aEM/\aEMs=0.9936\pm 0.0042$ and $\me/\mes=0.977^{+0.056}_{-0.071}$, which is in good agreement with the result given in \citet{Planck2015var_alp} for this case. The improvement is mainly due to better polarization information.

As for the analysis of $\me$, we can see that CMB data alone tends towards low values of $H_0$ and $\me/\mes$. This bias is removed when adding BAO, for which we find $\aEM/\aEMs=0.9989\pm 0.0026$, $\me/\mes=1.0056\pm 0.0080$ and $H_0=68.1\pm 1.3$. These numbers are consistent with the standard {\it Planck} 2015 cosmology (see Table~\ref{table:param_me}).

\vspace{-3mm}
\subsection{Constraining the redshift dependence of $\aEM$ and $\me$}
\label{sec:constraint_alp_me_p}
Next, we consider redshift-dependent variations of $\aEM$, using the parametrization in Eq.~\eqref{eq:powerlaw} with $\aEM(z_0)=\aEMs$. The constraints for this case are shown in third column of Table~\ref{table:param}. Since varying $p$ mainly affects the tilt of the CMB power spectra, degeneracies with $n_{\rm s}$ and $\Omega_{\rm b} h^2$ are expected. Indeed, we find the errors of these parameters to be slightly increased, while all other parameters are basically unaffected (see Fig.~\ref{fig:contour}). In particular, the $\theta_{\rm MC}$ contours still mimic the {\it Planck} 2015 contours without varying $\aEM$ as shown by the red and dotted black line in Fig.~\ref{fig:contour}. This already indicates that the individual effects of variations of $\aEM(z_0)$ and $p$ should be separable. When varying both parameters independently, we obtain the constraints indicated by the last column in Table~\ref{table:param}. Albeit slightly weakened, we can independently constrain $\aEM(z_0)$ and $p$. 

Carrying out a similar analysis for $\me$, setting $\me(z_0)=\mes$ we find $p=0.0006\pm 0.0044$ for {\it Planck} 2015 $TTTEEE$+lowP+lensing data. This is roughly $2$ times weaker than for $\aEM$, consistent with naive analysis of the free electron fraction scaling around $z\simeq 1100$ (see Fig.~\ref{fig:compare}). Also varying $\me(z_0)$, we obtain $\me/\mes=0.960^{+0.046}_{-0.071}$ and $p=0.0012^{+0.0047}_{-0.0042}$. When adding BAO data, this improves to $\me/\mes=1.0023\pm 0.0074$ and $p=0.0007\pm 0.0043$, again with no significant biases in the standard parameters with respect to the {\it Planck} 2015 cosmology remaining.

\vspace{-0mm}
\section{Conclusions}
\label{sec:conc} 
Current observations provide us with very precise cosmological datasets, that allow us to ask detailed questions about the conditions of the Universe around the recombination epoch.
In this paper, we analyzed the different effects on the recombination problem when varying $\aEM$ and $\me$. We explained the modifications to the recombination codes, {\tt Recfast++} and {\tt CosmoRec}, that are required to vary these constants in an easy and efficient way. In particular, we developed an improved correction function treatment for {\tt Recfast++} (Sect.~\ref{sec:mod_Recfast++}), which allows us to accurately represent the full computation of {\tt CosmoRec} (cf. Fig.~\ref{fig:compare}). We find that the remaining differences between the two recombination codes can in principle be neglected at the current level of precision.

For constant $|\dalpha|\lesssim 1\%$, we find a total effect on the ionization history of $\Delta X_{\rm e}/X_{\rm e}\simeq - 27 \times \dalpha$ at $z\simeq 10^3$ (see Fig.~\ref{fig:cont}). This is dominated by the required rescaling of the ionization potential in the equilibrium Boltzmann factors. Other corrections related to $A_{2\gamma}$, $\alpha_{\rm rec}$ and $\beta_{\rm phot}$ contribute at the $\sim 10\%$ level to this net effect. We also find that when varying $\aEM$ and $\me$ the associated direct changes to the recombination history caused by scaling the Thomson scattering cross section are negligible (see Fig.~\ref{fig:cont}). We still include this correction in our analysis for consistency.

When varying $\me$, we find the net effect on $X_{\rm e}$ around $z\simeq 10^3$ to be comparable to that of varying $\aEM$ for $\dme \approx 2.5 \times \dalpha$. \changeJ{The} net change of $X_{\rm e}$ in the freeze-out tail is smaller than for $\aEM$ (see Fig.~\ref{fig:cont_me} and \ref{fig:compare}), an effect that is related to the different scaling of $\alpha_{\rm rec}$ and $\beta_{\rm phot}$ with $\aEM$ and $\me$ (see Sect.~\ref{sec:effect_Xe}).

We also include explicit redshift-dependent variations of $\aEM$. This has a very different effect on the ionization history around recombination. Instead of shifting the recombination redshift during hydrogen recombination, the ionization history is stretch/compressed differentially, depending on the chosen parameters (see Fig.~\ref{fig:power_law}). This has a distinct effect on the CMB anisotropies that can be separated from the one for constant variations.

The propagation of the modifications in the recombination dynamics through to the Thomson visibility function and CMB anisotropies is also illustrated (see Sect.~\ref{sec:CL_results}). For constant $\dalpha$, our results are in agreement with previous analyses \citep[e.g.,][]{Kaplinghat1999, Battye2001}. We find that the changes to the CMB temperature power spectrum caused by variation of $\aEM$ and $\me$ are practically degenerate when $\dme \approx (2-3)\times \dalpha\simeq 10^{-3}$ (see Fig.~\ref{fig:talpha}). However, combined constraints on $\aEM$ and $\me$ are obtained in a regime in which higher order terms especially for $\me$ become relevant (Sect.~\ref{sec:alpha_me_sim}), making them again distinguishable. Changes caused by directly varying the CMB monopole temperature, $T_0$, should in principle be distinguishable due to the ISW effects (see Sect.~\ref{sec:T0_degen} and Fig.~\ref{fig:talpha}), although we do not explore this possibility in more detail here.

We also illustrate the effect of redshift-dependent changes. Instead of shifting the maximum of the Thomson visibility function (cf. Fig.~\ref{fig:visi_alpha}), a power-law variation of $\aEM$ with redshift (see Eq.~\ref{eq:powerlaw}) causes a change in the width of the visibility function (see Fig.~\ref{fig:visi_power}). This primarily modifies the blurring of CMB anisotropies (compare Fig.~\ref{fig:alpha_cl} and \ref{fig:power_cl}) and can thus be distinguished.

In Sect.~\ref{sec:constraints}, we present our constraints on different cases using {\it Planck} 2015 data. Our results (see Table~\ref{table:param} and \ref{table:param_me}) for constant $\dalpha$ and $\dme$ are consistent with those given in \citet{Planck2015var_alp}. We obtain the updated individual constraints $\aEM/\aEMs=0.9993\pm 0.0025$ and $\me/\mes=0.961^{+0.046}_{-0.072}$ using {\it Planck} 2015 data alone. Also adding BAO data, we find $\aEM/\aEMs=0.9997\pm 0.0023$ and $\me/\mes= 1.0039 \pm 0.0074$. When varying $\me$, the addition of BAO data removes the bias in $H_0$ towards low values, making the results consistent with the standard {\it Planck} cosmology (Table~\ref{table:param_me}). Simultaneous constraints when varying both $\aEM$ and $\me$ are presented in Sect.~\ref{sec:alpha_me_sim}.

Although we show that the effect of rescaling $\sigT$ for the computation of the Thomson visibility function is quite small (cf., Fig.~\ref{fig:visi_alpha} and \ref{fig:visi_power}), this effect should be included in the analysis \citep[see also][]{Planck2015var_alp}. For models with varying $\aEM$, we find that this effect improves the constraint by $\simeq 30\%$. The change is more dramatic when  varying $\me$. Here, we find that neglecting the rescaling of $\sigT$ leads to a significant underestimation of the error (a factor $\gtrsim 5$) unless BAO data is added. This is due to enhanced geometric degeneracies caused by the scaling of $\sigT$ in the visibility function calculation (see Sect.~\ref{sec:alpha_me}).

Allowing for power-law redshift dependence of $\aEM$ around $z_0=1100$ with $\aEM(z_0)=\aEMs$, we find the new constraint, $p=0.0008\pm 0.0025$, on the power-law index. When varying both $\aEM(z_0)$ and $p$, we obtain $\aEM(z_0)/\aEMs = 0.9998\pm 0.0036$ and $p = 0.0006\pm 0.0036$ (see Table~\ref{table:param}). Similarly, for $\me$ we find $p=0.0006\pm 0.0044$ (CMB only) assuming $\me(z_0)=\mes$. Varying both $\me(z_0)$ and $p$ we obtain $\me/\mes=1.0023\pm 0.0074$ and $p=0.0007\pm 0.0043$ when also adding BAO data (see Sect~\ref{sec:constraint_alp_me_p}).
All these results are fully consistent with the standard values, highlighting the impressive precision, stability and consistency of the data with respect to non-standard extensions. This also suggests that a wider class of varying fundamental constant models can in principle be probed using the CMB, possibly with more complex redshift-dependence (e.g., phase transition, spikes or higher order temporal curvature).

Modified recombination physics can also be investigated using CMB spectral distortions. For the future, we aim to continue this study with the cosmological recombination radiation \citep[e.g.,][]{Jose2006, Chluba2006b, Sunyaev2009}. Modeling these variations in {\tt CosmoSpec} \citep{Chluba2016CosmoSpec} will enlighten us on how the fundamental constants change the recombination spectrum and provide us with another dataset for constraints. This could allow us to alleviate existing parameter degeneracies and further deepen our understanding of the recombination epoch, allowing us to confront clear theoretical predictions with direct observational evidence. This might also open the possibility to probe the redshift-dependence of the fundamental constants at even earlier phases through the  individual effects on hydrogen and helium recombination (e.g., see Fig.~\ref{fig:power_law} for the effect on $X_{\rm e}$), which would remain inaccessible otherwise. One could furthermore refine constraints on spatial variations of fundamental constants. We look forward to exploring these opportunities.

\small 
\vspace{-0mm}
\section*{Acknowledgements}
We cordially thank the referee for their comments on the paper and the suggestions to more carefully consider the effect of $\sigT$ and extend our analysis for $\me$.
We also thank Marcos Iba\~{n}ez from IAC for his discussion surrounding {\tt CosmoMC} and how to optimise the sampling, given an increase in parameters. 
We extend thanks to Richard Battye, Carlos Martins and James Rich for discussion pertaining to previous studies of fundamental constant variations. 
We would also like to thank the Jodrell Bank Centre for Astrophysics, University of Manchester for the use of the {\sc Fornax} cluster as a tool to help reduce convergence runtimes.
LH is funded by the Royal Society through grant RG140523.
JC is supported by the Royal Society as a Royal Society University Research Fellow at the University of Manchester, UK.

\small 
\bibliographystyle{mn2e}
\bibliography{Lit}

\bsp	
\label{lastpage}
\end{document}